\documentclass[onecolumn,showpacs,preprintnumbers,amsmath,amssymb]{revtex4}
\newcommand{\boxh}{\hat{\mbox{\kern-.0em\lower.3ex\hbox{$\Box$}}}}

\newcommand{\be}{\begin{equation}}
\newcommand{\ee}{\end{equation}}
\newcommand{\bea}{\begin{eqnarray}}
\newcommand{\eea}{\end{eqnarray}}
\newcommand{\bean}{\begin{eqnarray*}}
\newcommand{\eean}{\end{eqnarray*}}
\newcommand{\bdm}{\begin{displaymath}}
\newcommand{\edm}{\end{displaymath}}

\newcommand{\nn}{\mbox{} \nonumber \\ \mbox{}}
\newcommand{\ba}{\begin{eqnarray}}
\newcommand{\ea}{\end{eqnarray}}

\newcommand{\lsim}{\stackrel{\scriptstyle <}{\scriptstyle \sim}}

\newcommand{\gbar}{\overline{\gbar}}
\usepackage{graphicx}% Include figure files
\usepackage{dcolumn}% Align table columns on decimal point
\usepackage{bm}% bold math

\begin{document}

\title{Force-Free Magnetohydrodynamic Waves: \\  Non-Linear
Interactions and Effects of Strong Gravity}

\author{Parker Troischt}
\affiliation{Department of Physics and Astronomy, University of
North Carolina, Chapel Hill, NC 27599}
 \email{troischt@physics.unc.edu}
\author{Christopher Thompson}
\affiliation{CITA, 60 St. George St., Toronto, ON, M5S 3H8}
 \email{thompson@cita.utoronto.ca}

\begin{abstract}
The propagation and non-linear interactions of magnetohydrodynamic waves 
are considered in the force-free limit, where the inertia of the conducting
matter which enforces the MHD condition ${\bf E}\cdot{\bf B} = 0$
can be neglected in comparison with the inertia of the
electromagnetic field.  By extending the analysis beyond the 
WKB approximation, we are able to study the non-linearities induced by a
gravitational field.  We treat
the perturbed electromagnetic field as a fluid of infinite conductivity.
We calculate the scattering of a torsional (Alfv\'en) wave by a gravitational
potential, and demonstrate a nonlinear coupling
with a compressive (fast) wave which is second order
in the amplitude of the Alfv\'en wave.  
In a cylindrically symmetric spacetime with slow rotation, the
coupling is second order in $g_{t\phi}$ and first order in
the amplitude of the wave.  
We also give a fresh analysis of the
non-linear interactions between compressive and torsional waves
in Minkowski space, with a focus on the relative strengths of
their three- and four-mode interactions.   In contrast with
non-relativistic magnetofluids, the effects of compression are
always present.  In the case of colliding fast waves, a net displacement 
of the field lines across (at least) one of the colliding wavepackets 
is shown to have a strong effect on the outgoing waveform, and to
have a qualitatively different interpretation than was previously
suggested for colliding Alfv\'en waves.  Finally, we show how spacetime 
curvature modifies the collision between two torsional waves, in both 
the weak- and strong-field regimes.
\end{abstract}

\pacs{Valid PACS appear here}% PACS, the Physics and Astronomy
\maketitle
\centerline{Submitted to PRD 11 May 2004; accepted 18 August 2004}% It is always \today, today,

%\begin{enumerate}

\section{Introduction}

	Magnetohydrodynamics can be studied in the ultra-relativistic
limit, where the energy density in the conducting matter which 
enforces ${\bf E}\cdot{\bf B} = 0$ is much less
than in the electromagnetic field itself.  The Euler equation reduces
to the force-free equation $J^\mu F_{\mu\nu} = 0$, which
provides a useful starting point for investigating some aspects
of the dynamics of accreting and outflowing matter around compact
stars, and classical gamma-ray bursts 
(e.g. \cite{BZ77}, \cite{OK97}, \cite{TB98}, \cite{LB03}).
In this regime a uniform magnetofluid supports two modes:
incompressible torsional waves analogous to Alfv\'en waves in 
non-relativistic magnetohydrodynamics (MHD), 
and a compressible wave analogous to the fast mode.  
Both of these modes have exact non-linear solutions in a uniform
magnetic field in Minkowski space, which means in turn that there 
is no spontaneous decay of one type of mode into the other. 

This fundamental property of a relativistic magnetofluid changes in
the presence of a strong gravitational field.   We focus in this
paper on the effect of spacetime curvature on the interactions between
torsional and compressive MHD modes.  There is a non-linear
interaction if the background spacetime is static, which appears at
linear order if it is slowly rotating.  We analyze the effect first
in the weak-field regime, and consider both spherically and
cylindrically symmetric gravitational fields.   As a particular
example where strong-field effects are important, we also consider
the case of a uniform magnetofluid immersed in the spacetime of
a black string.  

Our main goal here is to explore the fundamentals of the interactions
between MHD waves and gravitational fields.  Although the scattering
of gravitational waves, scalar waves, and vaccum electromagnetic
waves have long been studied in black hole spacetimes \cite{Chandra},
analogous work on force-free magnetohydrodynamics has focused
on the short-wavelength limit \cite{OK97}.
There are two well-established astrophysical contexts in which
the dynamics may be well described, in a first approximation,
by a relativistic force-free fluid:  the magnetosphere of 
a bursting magnetar (e.g. \cite{TD95}); and the polar regions of
an accreting black hole \cite{BZ77}.  We restrict ourselves,
in this paper, to the simplest case where the background
magnetofluid does not maintain a large-scale current
(as it does surrounding a Kerr black hole).

In a dynamic situation, a fluid formulation of 
force-free magnetohydrodynamics provides a physically transparent
set of variables.  Two such formalisms have been developed, by Achterberg
\cite{AT83} and Thompson \& Blaes \cite{TB98}.  These are,
in fact, dual versions of each other, and are connected
by interchanging the dynamical role of the Lagrangian fluid coordinates
and the Eulerian coordinates of the background spacetime.  In reviewing
these formalisms, we generalize the second to
an arbitrary curvilinear coordinate system.

Another of our goals is to extend the discussion of non-linear mode 
interactions in force-free MHD, begun in \cite{TB98}.  We examine in 
more detail how obliquely propagating Alfv\'en modes and fast modes will
perturb each other, including the effects of spacetime curvature.  The
effect of field line wandering on mode collisions has received 
considerable attention in recent years, in the case of 
non-relativistic and incompressible Alfv\'en turbulence.  We examine
its effect on the collisions between compressive (fast) modes in
a relativistic magnetofluid in flat Minkowski space.  In contrast
with the case of colliding Alfv\'en waves, the effect of a
net field-line displacement cannot be expressed in kinematic terms
through a zero-frequency component of one of the colliding waves.

The sections are organized as follows.  In Section II, 
we given an overview the electric and magnetic lagrangian fluid formulations
of force-free MHD.  As a toy example, we
construct the solution for the uniform magnetic field in a spherically
symmetric spacetime (with non-vanishing Ricci tensor).  In Section III
we review the normal modes of a uniform, force-free magnetofluid.
Section IV is devoted to the scattering of a torsional wave in
a curved spacetime:  we first consider the propagation of such
a wave in a general, cylindrically symmetric spacetime;  and then
consider the scattering of such a wave in a shallow, spherical gravitational
potential.  Section V generalizes the results of the previous
section to the case of two colliding, axisymmetric torsional modes
in a gravitational field.
The final Section VI considers the more general case of collisions
between non-axisymmetric MHD modes in Minkowski space, focusing
on the three-mode couplings of two colliding Alfv\'en and fast waves.

\section{Lagrangian Formulations of Ultrarelativistic MHD}

The equations describing the behavior of ideal fluids coupled to
electromagnetic fields in curved spacetimes are inherently very complicated.
A significant simplification is obtained by assuming that the spacetime
metric is fixed; i.e., that the mass-energy of the fluid can be neglected
in comparison with that of the source of gravity.  In this paper
we make a second simplification:  the role of the charged matter fields
is restricted to a source of electric currents which cancel off
the electric field in the fluid rest frame.  The mass-energy of these
matter fields is, in other words, negligible in comparison with that
of the electromagnetic field.  

We begin by summarizing the two Lagrangian fluid formulations of
relativistic, force-free MHD.

\subsection{MHD equations in the force-free limit}

The covariant Maxwell Equations are
\be
4 \pi J^{\mu} = {\nabla}_{\nu} F^{\mu \nu} = {1\over \sqrt{-g}}
\partial_\nu\left(\sqrt{-g}g^{\mu\alpha}g^{\nu\beta}F_{\alpha\beta}\right)
\ee
and
\be
{\nabla}_{\alpha} F_{\mu \nu} + {\nabla}_{\nu} F_{\alpha \mu}
+ {\nabla}_{\mu} F_{\nu \alpha} = 
\partial_\alpha F_{\mu\nu} + \partial_\nu F_{\alpha\mu} +
\partial_\mu F_{\nu\alpha} = 0.
\ee
When the only force acting on the charged matter in a local
inertial frame is the Lorentz force, the dynamics of the electromagnetic
field is described by the force-free equation
\be
J^\mu F_{\mu\nu} = {1\over 4\pi}
{\nabla}_{\alpha}F^{\mu\alpha} F_{\mu\nu} = 0
\ee
in combination with
\be
F_{\mu\nu}\widetilde F^{\mu\nu} = 0.
\ee
Here
\be
\widetilde F^{\mu\nu} = {1\over 2\sqrt{-g}}\varepsilon^{\mu\nu\alpha\beta}
F_{\alpha\beta}
\ee
is the dual of the electromagnetic field tensor, expressed in terms
of the antisymmetric symbol with unit norm $|\varepsilon^{\mu\nu\alpha\beta}|
= 1$.

\subsection{Electric lagrangian variables}

Achterberg \cite{AT83} developed
a variational principle for relativistic magnetohydrodynamics in
which the Lagrangian coordinates $x_0^\mu$ of the fluid are the field 
variables, moving in a fixed Eulerian space $x^\mu$.
In the ideal MHD limit, where the electric field vanishes in the rest
frame of the fluid, the Lie derivative of the Faraday two-form
vanishes, ${\pounds}_u F = 0$, along the direction $u$ of fluid motion.
One thereby obtains a simple relation
between the background (reference) state and the dynamic state of the
magnetofluid.  Solving for the $x_0^\mu$ allows one to
obtain all relevant information about the system.  

The Faraday two-form is related to its background counterpart by
\be\label{transf}
F_{\mu \nu} = \frac{\partial {x_0}^\alpha}{\partial {x^{\mu}}}
\frac{\partial {x_0}^\beta}{\partial {x^{\nu}}}{F^{(0)}_{\alpha\beta}}.
\ee
In the force-free limit, the action is
\be
S = -\int d^4x  \frac {1}{4} \sqrt{-g} g^{\mu\rho}
g^{\nu\sigma}F_{\mu \nu}F_{\rho \sigma}.
\ee
By extremizing this action, the equation of motion
\be
J^{\mu}F_{\mu \nu} = 0
\ee
can be obtained.  Written in terms of the coordinate fields this is
\be
\frac{\partial x_0^\alpha}{\partial x^\mu}
F^{(0)}_{\alpha\nu}
\frac{\partial}{\partial {x^\rho}}
\left(\sqrt {-g}g^{\rho\sigma}g^{\mu\tau} 
\frac{\partial x_0^\gamma}{\partial {x^\sigma}}
\frac{\partial x_0^\delta}{\partial x^\tau}
F^{(0)}_{\gamma\delta}\right) = 0.
\ee
Here, we have eliminated a factor of $\partial x_0^\beta/\partial x^\nu$ from
$F_{\mu\nu}$.
The Bianchi identities are automatically satisfied as long as they are
satisfied by the background field.
Once the equations for these coordinate fields have been solved, the
electromagnetic field is obtained from Eq. (\ref{transf}).

\subsection{Magnetic lagrangian variables}

Two new MHD formalisms in the extreme relativistic limit were recently
developed by Thompson and Blaes \cite{TB98}.  The first of these is related
by a duality transformation to the action presented by Achterberg, and is
of interest here. In this formalism, the dual tensor $\widetilde{F}$ (rather
than $F$) plays a central role and the dynamical fields are the perturbed
coordinates $x^{\mu}$ (rather than the $x_0^{\mu}$). In this
section, we generalize the treatment of \cite{TB98} to include gravity.

Once the fields $x^\mu$ are solved, the electromagnetic fields can be found 
from the transformation
\be\label{magvar}
\widetilde F^{\mu\nu} = \frac{\sqrt{-g(x_0)}}{J_4\sqrt{-g(x)}}
\frac{\partial x^\mu}{\partial x_0^\alpha}
\frac{\partial x^\nu}{\partial x_0^\beta}\widetilde F_0^{\alpha\beta}.
\ee
Here
\be
J_4 = \det\left({\partial x^\mu\over \partial x_0^\alpha}\right)
\ee
is the Jacobian of the transformation $x_0^\alpha \rightarrow x^\mu$
from Lagrangian to Eulerian variables, and the factor of $\sqrt{g_0/g}$
is necessary in a curvilinear coordinate system.  
We show in Appendix \ref{appone} that the force-free equation
$J^\mu F_{\mu\nu} = 0$ may be derived from the action
\be
S' = \int d^4x_0 L^\prime = 
\int d^4x \frac {1}{4} \sqrt{-g} g_{\mu\rho} g_{\nu\sigma}
{\widetilde F}^{\mu\nu}{\widetilde F}^{\rho\sigma} =
\int d^4x_0 \frac{J_4}{4} \sqrt{-g} g_{\mu\rho} g_{\nu\sigma}
{\widetilde F}^{\mu\nu} {\widetilde F}^{\rho\sigma}.
\ee
Along the way, we obtain the equation of motion
\be\label{ffc}
\widetilde F^{\delta\varepsilon}{\partial\over\partial x^\varepsilon}
\left[g_{\beta\delta}g_{\gamma\mu}\widetilde F^{\beta\gamma}\right]
- {1\over 2}\widetilde F^{\delta\varepsilon}
{\partial\over\partial x^\mu}\left[g_{\beta\delta}g_{\gamma\varepsilon}
\widetilde F^{\beta\gamma}\right] = 0.
\ee

In cases where the background magnetic field asymptotes to a constant
at large distances, we can use the simple background 
\be\widetilde F_0^{\mu\nu} = 
B_0(\delta^\mu_0\delta^\nu_3-\delta^\mu_3\delta^\nu_0);\;\;\;\;\;\;
\sqrt{-g_0} = 1,
\ee
and define
proper time $\tau \equiv t_0$ and distance $\sigma \equiv z_0$.  The
equation of motion (\ref{ffc}) then takes the form
\be\label{ffd}
{\partial x^\varepsilon\over\partial x_0^\alpha}
\left(
{\partial x^\delta\over\partial\tau}{\partial\over\partial\sigma} - 
      {\partial x^\delta\over\partial\sigma}{\partial\over\partial\tau}
\right)\Sigma_{\delta\varepsilon}
-\left({\partial x^\delta\over\partial\tau}
{\partial x^\varepsilon\over\partial\sigma}
% - {\partial x^\varepsilon\over\partial\tau}
%{\partial x^\delta\over\partial\sigma}
\right){\partial\over\partial x_0^\alpha}\Sigma_{\delta\varepsilon} = 0,
\ee
where
\be
\Sigma_{\mu\nu} \equiv {g_{\mu\alpha}g_{\nu\beta}\over
\sqrt{-g}J_4}\left({\partial x^\alpha\over\partial \tau}
{\partial x^\beta\over\partial \sigma} - 
{\partial x^\beta\over\partial \tau}
{\partial x^\alpha\over\partial \sigma}\right) = -\Sigma_{\nu\mu}.
\ee 
Although Eq. (\ref{ffc}) has four components, it
is straightforward to check that the longitudinal components
$\mu = 0, 3$ vanish.  Physically, this is equivalent to the statement
that MHD waves have two independent transverse polarizations.  
Notice also that it is possible to simplify Eq. (\ref{ffc}) further
by choosing a time slicing in which
\be
\sqrt{-g}J_4 = 1.
\ee
The cost of this gauge choice is that one can no longer identify
the background time coordinate $t$ with the proper time $\tau$ of
the magnetofluid.

\subsection{Uniform magnetic field: Spherically symmetric spacetime}
\label{magsphere}

There is a simple prescription for constructing a uniform magnetic
field ($J^\phi = 0$) in a vacuum spacetime with an axial killing vector
$\xi^\mu = {g^\mu}_\phi$.  One identifies \cite{W74}  
\be
A_\mu = {1\over 2}B_0\xi_\mu = {1\over 2}B_0 g_{\phi\mu}.
\ee
Thus $F_{\mu\nu}$ takes exactly the same form outside a black hole
of vanishing charge and spin, in the standard Schwarzschild 
coordinates ($g_{\phi\phi} = r^2\sin^2\theta$), as it does
in Minkowski space:
\be\label{fminksph}
F_{r\phi} = B_0 r \sin^2\theta;\;\;\;\;\;\;F_{\theta\phi} = 
B_0r^2\sin\theta\cos\theta.
\ee

More generally this method is of limited utility, since it depends on the 
vanishing of the Ricci tensor of the background spacetime.  
To give a further illustration of the
lagrangian fluid method, we show how it may be used to construct a uniform
magnetic field in a general spherical spacetime.  The metric may be written
as
\be
ds^2 = g_{tt}dt^2 + g_{rr}dr^2 + r^2(d\theta^2+\sin^2\theta d\phi^2).
\ee
We imagine that the currents
supporting the magnetic field are flowing far outside the 
self-gravitating mass, which itself is not electrically 
conducting.  Then
\be 
F_{\mu\nu} = (\partial_\mu R_0 \partial_\nu\phi_0 - 
\partial_\mu\phi_0\partial_\nu R_0)F^0_{R\phi}
= {1\over 2}(\partial_\mu R_0^2 \partial_\nu\phi - 
\partial_\mu\phi_0\partial_\nu R_0^2)B_0.
\ee
Here $R_0 = r_0\sin\theta_0$ is the cylindrical radius and
$F^0_{R\phi} = B_0R_0$ is the field strength that would 
be supported by the same currents in flat space.  The vanishing of 
\be
4\pi\sqrt{-g}J^\phi = \partial_r\left(\sqrt{-g}g^{\phi\phi}g^{rr}
F_{\phi r}\right) + 
\partial_\theta\left(\sqrt{-g}g^{\phi\phi}g^{\theta\theta}
F_{\phi \theta}\right)
\ee
combined with the separation of variables
\be
r_0^2\sin^2\theta_0 = {\cal R}(r)\,{\cal T}(\theta)
\ee
gives the coupled ordinary differential equations
\be
{r^2\over {\cal R}\sqrt{-g_{tt}g_{rr}}}{d\over dr}
\left(\sqrt{-{g_{tt}\over g_{rr}}} {d{\cal R}\over dr}\right)
= - {\sin\theta\over{\cal T}}{d\over d\theta}
\left({1\over\sin\theta}{d{\cal T}\over d\theta}\right).
\ee
Because the equation for ${\cal T}$ is the same as in flat space, we may
write
\be 
\sin\theta_0 = \sin\theta
\ee
and
\be\label{fpert}
F_{r\phi} = {1\over 2}B_0 {d{\cal R}\over dr} \sin^2\theta;\;\;\;\;\;\;
F_{\theta\phi} = B_0{\cal R}\sin\theta\cos\theta,
\ee
where ${\cal R}$ is a solution to 
\be\label{Rfeq}
r^2{d\over dr}\left(\sqrt{-g_{tt}\over g_{rr}}
{d{\cal R}\over dr}\right) - 2\sqrt{-g_{tt}g_{rr}}{\cal R} = 0.
\ee
If the Ricci tensor of the background spacetime vanishes at large 
radius, then the appropriate solution to (\ref{Rfeq}) must match 
onto ${\cal R}(r) = r^2$ at large $r$ (where $-g_{tt} = g_{rr}^{-1} 
= 1-2GM/r$).

\subsection{Uniform electromagnetic field: Cylindrically symmetric spacetime}

We will be concerned with the propagation of MHD waves in
cylindrically symmetric geometries of the following form
\be
{ds}^2 = g_{tt} {dt}^2 + g_{rr}{dr}^2 + g_{\phi \phi}{d{\phi}}^2 + g_{zz}{dz}^2
+ 2g_{t \phi} dt d{\phi}.
\ee
When $g_{t\phi}\neq 0$, the frame dragging of the background $z$-magnetic
field generates a radial electric field.  Thus,
the background electromagnetic field is a stationary, axisymmetric, 
and $z$-independent solution to the equation $J^\mu = 0$ with
non-vanishing components $F^{(0)}_{r\phi}$ and $F^{(0)}_{rt}$,
\be
\partial_{r}\left(\sqrt{-g}g^{tt}g^{rr}F^{(0)}_{tr} +
\sqrt{-g}g^{t \phi}g^{rr}F^{(0)}_{\phi r}\right) = 0;
\ee
\be\label{jphiback}
\partial_{r}\left(\sqrt{-g}g^{\phi \phi}g^{rr}F^{(0)}_{\phi r} + 
\sqrt{-g}g^{\phi t}g^{rr} F^{(0)}_{tr}\right) = 0.
\ee
Since the electric field vanishes in the static limit,
one has
\be\label{eback}
 F^{(0)}_{rt} = -{g_0^{t \phi}\over g_0^{tt}} F^{(0)}_{r \phi} 
\ee
and 
\be\label{bback}
F^{(0)}_{r \phi} = \frac{B_0}{\sqrt{-g_0} g_0^{rr} [g_0^{\phi \phi} -
(g_0^{t \phi})^2 / g_0^{tt}]} = \frac{g_{0\,rr}g_{0\,\phi\phi}}
{\sqrt{-g_0}}B_0.
\ee
[Here $g_0 \equiv g(x_0)$.]

We will investigate perturbations to the fields, subject to the constraint
$F^{\mu\nu}\widetilde F_{\mu\nu} = 0$, in spacetimes with varying degrees 
of curvature.   It is useful first to consider the case of
a nearly flat and non-rotating cylindrical spacetime.  In general, 
static spacetimes which are symmetric about the z-axis 
may be written in following form \cite{S60}
\be\label{weak0}
ds^2 = - e^{2 \lambda}dt^2 + e^{2(\nu - \lambda)}(dr^2 + dz^2) 
+ r^2e^{-2 \lambda}d{\phi}^2.
\ee
The function $\lambda$ satisfies Laplace's equation in cylindrical coordinates,
and $\nu$ is second order in $\lambda$.  In the weak field limit 
this metric reduces to
\be\label{weak}
ds^2 = - (1 + 2 \lambda)dt^2 + (1 - 2 \lambda)(dr^2 + dz^2) + 
r^2(1 - 2 \lambda)d{\phi}^2,
\ee
and 
\be
\lambda = -G \int d^3x' \frac {\rho(x')}{|x - x'|}
\ee 
becomes the usual Newtonian potential.  
The background fields (\ref{bback}) and (\ref{eback}) now reduce to
\be
F^{(0)}_{r \phi} =  (1 - 2\lambda)\,B_0r;\;\;\;\;\;\;
F^{(0)}_{rt} = 0.
\ee

A cylindrically symmetric spacetime with some attributes of
$3+1$-dimensional black holes can be constructed from the
product of $2+1$-dimensional black hole (sitting in the $r$-$\phi$ plane)
and an infinite line segment along the $z$-axis.
The $2+1$ black hole solution was originally found by Banados,
Teitelboim and Zanelli \cite{BTZ92}; its properties are reviewed by
Carlip \cite{CL95}.  The key element in its construction
is the introduction of a negative cosmological constant
$\Lambda = - \frac {1} {\ell^2}$, so that the spacetime asymptotes 
to anti-de Sitter space at large radius.  The BTZ metric is
\be\label{dsbtzj}
{ds}^2 = - \left(-M + \frac{r^2} {\ell^2} + \frac {J^2}{4r^2}\right) {dt}^2 +
{\left(-M + \frac{r^2} {\ell^2} + \frac {J^2}{4r^2}\right)}^{-1}{dr}^2
+ r^2 \left({d{\phi}} - \frac {J}{2r^2}dt\right)^2 + {dz}^2,
\ee
or equivalently
\be
{ds}^2 = - (ZR) {dt}^2 +
(ZR)^{-1}{dr}^2
+ r^2 \left({d{\phi}} - \frac {J}{2r^2}dt\right)^2 + {dz}^2,
\ee
where
\be
Z \equiv -M + \frac{r^2} {\ell^2}
\ee
and
\be
R \equiv 1 - {g_{t \phi}^2\over g_{tt}g_{\phi\phi}}
= 1 + \frac {J^2}{4Zr^2}.
\ee
(The variable $M$, being a mass per unit length, is dimensionless
in units where $G = c = 1$.)
The background Maxwell fields in this spacetime are
\be\label{Fbrphi}
F^{(0)}_{r \phi} = \frac {B_0 r} {ZR};
\ee
\be\label{Fbrt}
F^{(0)}_{rt} = \frac{J}{2r^2} F^{(0)}_{r \phi} =
\left(\frac{J}{2r^2}\right) \frac {B_0 r} {ZR}.
\ee
These expressions reduce to 
\be
F^{(0)}_{r \phi} = {B_0r\over r^2/\ell^2-M};\;\;\;\;\;\;F^{(0)}_{rt} = 0
\ee
in the static limit ($J = 0$). Minkowski space is recovered by 
taking  $M \rightarrow -1$ and 
$\ell^2 \rightarrow \infty$.

\section{Force-Free MHD Waves in Curved Spacetimes}

\subsection{MHD modes:  Force-free limit}

We first review the normal modes of a uniform magnetofluid in
flat Minkowski space, before considering their interactions and
the effects of spacetime curvature.   The background
magnetic field is ${\bf B} = B_0\hat z$, or equivalently
$F_{r \phi} = B_0r$ in cylindrical coordinates.  The fluid supports
two distinct modes in the force-free limit:
the fast mode, which involves a compressive disturbance 
and has an isotropic dispersion relation $\omega = |{\bf k}|$; 
and the Alfv\'en mode, which is incompressible and has a group
velocity directed along the background magnetic field,
$d\omega/dk_z = \pm 1$.  There is no slow mode.

The simplest way of seeing that the fluid supports only two normal
modes is to note that the perturbation to the Faraday tensor
\be
\delta F_{\mu \nu} = F_{\mu \nu} -  F^{(0)}_{\mu \nu},
\ee
can be expressed in terms of two variables $\delta r$ and $\delta\phi$.
To first order in the perturbation, we have
\be\label{dfrphi}
\delta F^{(1)}_{r \phi} =
\frac{\partial}{\partial r}\left(F^{(0)}_{r \phi} \delta r\right)
+{\partial\delta\phi\over\partial\phi}F^{(0)}_{r\phi};
\;\;\;\;\;\;\delta F^{(1)}_{t \phi} = \frac{\partial\delta r}{\partial t}
F^{(0)}_{r \phi};
\;\;\;\;\;\;\delta F^{(1)}_{z \phi} = \frac{\partial\delta r}{\partial z}
F^{(0)}_{r \phi},
\ee
and
\be\label{dfrphib}
\delta F^{(1)}_{tr} = -{\partial\delta\phi\over\partial t}F^{(0)}_{r\phi};
\;\;\;\;\;\;
\delta F^{(1)}_{zr} = -{\partial\delta\phi\over\partial z}F^{(0)}_{r\phi};
\;\;\;\;\;\;\delta F^{(1)}_{tz} = 0.
\ee
As expected, the component of the electric field parallel to ${\bf B}_0$
vanishes to linear order.
The absence of the slow mode can be related to the invariance of
$F^{(0)}_{\mu\nu}$ under reparameterizations of the $z$-coordinate.  
The full expressions for $\delta F_{\mu\nu}$, written in terms of
the Lagrangian fluid variables, are collected in Appendix A.

More generally, the Alfv\'en mode and fast mode involve perturbations
to both fluid coordinates.  These perturbations are related by the condition
of incompressibility
\be\label{alfcon}
{1\over r}{\partial(r\delta r)\over\partial r} + {\partial\delta\phi\over
\partial\phi} = 0\;\;\;\;\;\;(\rm Alfven)
\ee
in the case of the Alfv\'en mode; and the condition of vanishing
torsion
\be\label{fastcon}
{\partial\delta r\over\partial\phi}-{\partial(r^2\delta\phi)\over\partial r}
= 0\;\;\;\;\;\;(\rm Fast)
\ee
in the case of the fast mode.   The fluctuating current associated with 
the field perturbations (\ref{dfrphi}), (\ref{dfrphib}) is
\be\label{jtexp}
4\pi\sqrt{-g}J^{t(1)} = 
\partial_r\left(\sqrt{-g}g^{tt}g^{rr}\delta F^{(1)}_{tr}\right) +
\partial_\phi\left(\sqrt{-g}g^{tt}g^{\phi\phi}\delta F^{(1)}_{t\phi}\right);
\ee
\be\label{jzexp}
4\pi\sqrt{-g}J^{z(1)} =
\partial_r\left(\sqrt{-g}g^{zz}g^{rr}\delta F^{(1)}_{zr}\right) +
\partial_\phi\left(\sqrt{-g}g^{zz}g^{\phi\phi}\delta F^{(1)}_{z\phi}\right);
\ee
\be\label{jrexp}
4\pi\sqrt{-g}J^{r(1)} =
\partial_t\left(\sqrt{-g}g^{rr}g^{tt}\delta F^{(1)}_{rt}\right) +
\partial_z\left(\sqrt{-g}g^{rr}g^{zz}\delta F^{(1)}_{rz}\right) +
\partial_\phi\left(\sqrt{-g}g^{rr}g^{\phi\phi}\delta F^{(1)}_{r\phi}\right);
\ee
and
\be\label{jphiexp}
4\pi\sqrt{-g}J^{\phi(1)} =
\partial_r\left(\sqrt{-g}g^{\phi \phi}g^{rr}\delta F^{(1)}_{\phi r}\right) +
\partial_t\left(\sqrt{-g}g^{\phi \phi}g^{tt}\delta F^{(1)}_{\phi t}\right) +
\partial_z\left(\sqrt{-g}g^{\phi \phi}g^{zz}\delta F^{(1)}_{\phi z}\right).
\ee
Expressed in terms of the coordinate perturbations, and specializing
to Minkowski space, this becomes
\be
4\pi r J^{t(1)} = 
B_0\,\partial_t\left[\partial_r\left(r^2\delta\phi\right)
           -\partial_\phi\delta r\right];
\ee
\be
4\pi r J^{z(1)} = 
-B_0\,\partial_z\left[\partial_r\left(r^2\delta\phi\right)
           -\partial_\phi\delta r\right];
\ee
\be\label{jrgen}
4\pi r J^{r(1)} =
B_0r^2\left\{-\partial_t^2\delta\phi + \partial_z^2\delta\phi
+ {1\over r^2}\partial_\phi\left[{1\over r}\partial_r\left(r\delta r\right)
+\partial_\phi\delta\phi\right]\right\};
\ee
and
\be\label{jphigen}
4\pi r J^{\phi(1)} =
-B_0\left\{-\partial_t^2\delta r + \partial_z^2\delta r + 
\partial_r\left[{1\over r}\partial_r\left(r\delta r\right) +
\partial_\phi\delta\phi\right]\right\}.
\ee

The dynamical equations for the two normal modes are recovered 
by imposing the force-free condition.  To linear order one has
\be\label{fflin}
J^{r(1)}F^{(0)}_{r\phi} = 0 = J^{\phi(1)}F^{(0)}_{\phi r},
\ee
hence $J^{r(1)} = J^{\phi(1)} = 0$.
In the case of the Alfv\'en mode, applying the constraint
(\ref{alfcon}) gives the one-dimensional wave equation
\ba\label{alfwave}
-\partial_t^2\delta r + \partial_z^2\delta r &=& 0;\nn
-\partial_t^2\delta\phi + \partial_z^2\delta\phi &=& 0,
\ea
with a vanishing longitudinal field perturbation
\be
\delta F^{(1)}_{r\phi} = 0,
\ee
and (in general) a non-vanishing longitudinal current 
\be
J^{t(1)},\; J^{z(1)} \neq 0.
\ee
A force-free magnetofluid supports torsional waves of both helicities.
It should be noted that, in this limit, there is
no basic distinction between the two helicities, 
as there is a non-relativistic cold, magnetized plasma.

In the case of the fast mode, one instead finds that the current
vanishes entirely and the coordinate perturbations obey 
three-dimensional wave equations,
\ba\label{rmink}
-\partial_t^2\delta\phi + \partial_z^2\delta\phi +
{1\over r^3}\partial_r\left[r\partial_r\left(r^2\delta\phi\right)\right] +
{1\over r^2}\partial_\phi^2\delta\phi &=& 0;\nn
-\partial_t^2\delta r + \partial_z^2\delta r +
\partial_r\left[{1\over r}\partial_r\left(r\delta r\right)\right] +
{1\over r^2}\partial_\phi^2\delta r &=& {2\over r}\partial_\phi\delta\phi.
\ea
By applying the operator $r^{-1}\partial_r r$ to the first equation,
one sees that an axisymmetric field perturbation
\be
{\delta B^z\over B_0} = {\delta F^{(1)}_{\phi r}\over F^{(0)}_{\phi r}} =
{1\over r}{\partial\over\partial r}\left(r\delta r\right)
\ee
satisfies the (axisymmetric) cylindrical wave equation.  

Thus, the fast mode is equivalent to a vacuum electromagnetic
wave superposed on the uniform background magnetic field, but with its
polarization restricted to $\delta{\bf E}\cdot{\bf B}_0 = 0$.
This way of constructing the fast mode also makes it immediately
apparent that infinite plane waves (restricted to this single polarization
state) are another general class of solutions.  

Inspection of Eqs. (\ref{jrgen}) and (\ref{jphigen}) shows that another
projection,
\be\label{fastalt}
{1\over r^2}{\partial\delta r\over\partial\phi} - 
{\partial\delta\phi\over\partial r} = 0,
\ee
allows the equations of motion for $\delta r$ and $\delta\phi$ to
be entirely separated,
\ba\label{rmink2}
-\partial_t^2\delta\phi + \partial_z^2\delta\phi +
{1\over r^3}\partial_r\left(r^3\partial_r\delta\phi\right) +
{1\over r^2}\partial_\phi^2\delta\phi &=& 0;\nn
-\partial_t^2\delta r + \partial_z^2\delta r +
\partial_r\left[{1\over r}\partial_r\left(r\delta r\right)\right] +
{1\over r^2}\partial_\phi^2\delta r = 0.
\ea
At the same time, the longitudinal currents do not vanish,
\be\label{longcur}
J^{t(1)} = {B_0\over 2\pi}\partial_t\delta\phi;
\;\;\;\;\;\; J^{z(1)} = -{B_0\over 2\pi}\partial_z\delta\phi.
\ee

\subsection{Decomposition into normal modes in a curved spacetime}

We have seen that, even in Minkowski space, there is an ambiguity in
the definition of the fast mode.  Axially symmetric perturbations
decompose directly into torsional modes, which are supported only
by a perturbation $\delta\phi$, and compressive modes, which are
supported only by a perturbation $\delta r$.  In the more
general case of non-axisymmetric perturbations, one can still define
unambiguously an incompressible mode (\ref{alfcon}).  But one is
faced with a choice either of defining the fast mode to be 
current free through Eq. (\ref{fastcon}), in which case the equations of
motion (\ref{rmink}) for $\delta r$ and $\delta\phi$ do not entirely 
separate; or instead of using the alternative projection (\ref{fastalt})
which separates these equations but leaves a non-vanishing 
longitudinal current (\ref{longcur}).  

This ambiguity remains when one considers the case of a magnetofluid
in a gravitational field.  We focus here on a general static
($g_{t\phi} = 0$) and cylindrically symmetric spacetime.  In this 
case, the longitudinal currents (\ref{jtexp}) and (\ref{jzexp})
cannot both vanish, as is seen by writing them in terms of the
coordinate perturbations:
\be\label{jtcurve}
{4\pi\over B_0}\sqrt{-g}J^{t(1)} = {\partial\over\partial t}
\left[-{\partial\over\partial r}\left({g_{\phi\phi}\over g_{tt}}
\delta\phi\right) +
{g_{rr}\over g_{tt}}{\partial\delta r\over\partial \phi}\right];
\ee
and 
\be\label{jzcurve}
{4\pi\over B_0}\sqrt{-g}J^{z(1)} = {\partial\over\partial z}
\left[-{\partial\over\partial r}\left({g_{\phi\phi}\over g_{zz}}
\delta\phi\right) +
{g_{rr}\over g_{zz}}{\partial\delta r\over\partial \phi}\right].
\ee
In these expressions, $g_{tt}$ and $g_{zz}$ have different dependences
on radius $r$, and we have made use of the diagonality of the metric.  
We conclude that, to linear order in the field perturbation,
non-axisymmetric excitations of the fast mode are current-carrying.

It is, nonetheless, possible to choose a generalization of the
projection (\ref{fastalt}) which does allow the equations of
motion for $\delta r$ and $\delta\phi$ to decouple at linear order.  
From Eqs. (\ref{jtcurve}) and (\ref{jzcurve}), the obvious choice is
\be\label{fastaltb}
{g_{rr}\over g_{\phi\phi}}{\partial\delta r\over\partial\phi} -
{\partial\delta\phi\over\partial r} = 0.\;\;\;\;\;\;({\rm Fast})
\ee
Then Eq. (\ref{jphiexp}) becomes
\be
\left({1\over g_{tt}}\frac{\partial^2\delta r}{\partial t^2} + 
{1\over g_{zz}}\frac{\partial^2\delta r}{\partial z^2}\right)F^{(0)}_{r\phi} +
{1\over g_{\phi\phi}}{\partial^2\delta r\over\partial \phi^2}
F^{(0)}_{r\phi} +
{g_{\phi\phi}\over\sqrt{-g}}\frac{\partial}{\partial r}
\left[\frac{\sqrt{-g}}{g_{\phi \phi} g_{rr}} \frac{\partial}{\partial r}
\left(\delta r F^{(0)}_{r \phi}\right)\right] = 0,
\ee
or, upon substituting expression (\ref{bback}),
\be\label{waveeqa}
{1\over g_{tt}}\frac{{\partial}^2 \delta r}{\partial t^2} +
{1\over g_{zz}}{\partial^2\delta r\over\partial z^2} +
{1\over g_{\phi\phi}}{\partial^2\delta r\over\partial \phi^2} +
{1\over g_{rr}}\frac{\partial}{\partial r}
\left[\frac{\sqrt{-g}}{g_{\phi \phi} g_{rr}} \frac{\partial}{\partial r}
\left(\delta r {g_{rr}g_{\phi\phi}\over\sqrt{-g}}\right)\right] = 0.
\ee
The longitudinal currents then are
\be
{4\pi\over B_0}\sqrt{-g}J^{t(1)} = -{\partial\over\partial r}
\left({g_{\phi\phi}\over g_{tt}}\right){\partial\delta\phi\over\partial t};
\;\;\;\;\;\;
{4\pi\over B_0}\sqrt{-g}J^{z(1)} = -{\partial\over\partial r}
\left({g_{\phi\phi}\over g_{zz}}\right){\partial\delta\phi\over\partial z}.
\ee
The perturbation $\delta\phi$ can be recovered from $\delta r$
using the constraint (\ref{fastaltb}).  Similarly, the incompressible
Alfv\'en mode is defined by
\be
{\sqrt{-g}\over g_{rr}g_{\phi\phi}}{\partial\over\partial r}
\left({g_{rr}g_{\phi\phi}\over\sqrt{-g}}\delta r\right) +
{\partial\delta\phi\over\partial\phi} = 0,\;\;\;\;\;\;({\rm Alfven})
\ee
to linear order.

For example, the wave equation for a purely compressive disturbance
in a BTZ black string spacetime is
\be
- {1\over Z^2} \frac {{\partial}^2 \delta r}{\partial t^2}
+ {1\over Z}{\partial^2\delta r\over\partial z^2}
+ {1\over r^2 Z}{\partial^2\delta r\over\partial \phi^2}
+ \frac {\partial}{\partial r}
\left[\frac {Z}{r} \frac {\partial}{\partial r}
\left(\frac{r \delta r}{Z}\right)\right] = 0.
\ee
(As before, $Z(r) \equiv r^2/\ell^2-M$.)
When $\partial_z\delta r = \partial_\phi\delta r = 0$, this equation 
can also be re-expressed directly in terms of the field perturbation 
\be
{\delta B^z\over B^z} = {Z\over r}{\partial\over\partial r}
\left({r\delta r\over Z}\right)
\ee
as
\be\label{waveBTZ}
- \frac{\partial^2}{\partial t^2}
\left({\delta B^z\over B^z}\right)
+ {Z\over r}{\partial\over\partial r}
\left[Zr {\partial\over \partial r}
\left({\delta B^z\over B^z}\right)\right] = 0.
\ee
More generally, the field perturbations have some dependence on 
$z$, and it is more useful to work with the coordinate
perturbation.

\section{Scattering of Torsional Waves by Spacetime Curvature}

We now turn to an important, but subtle,
difference between the compressive (fast) and
torsional (Alfv\'en) modes in a curved spacetime.
We focus on a uniform magnetofluid in which the background current
vanishes.   Both modes
are exact {\it non-linear} solutions to the force-free equation
in Minkowski space.  This property is retained by a purely
radial oscillation ($\delta\phi = 0$), in any 
static cylindrically symmetric (or spherically symmetric) spacetime.
The resulting fast mode is equivalent to an electromagnetic 
wave polarized in the $\phi$-direction, with
$\delta{\bf  E}\cdot {\bf B}_0 = 0$.  Hence it can have arbitrary amplitude.  

An axisymmetric torsional wave is not, by contrast, a 
non-linear solution to $J^\mu F_{\mu\nu} = 0$, because the wave 
speed varies with radius.  We now show that, in a static spacetime, 
such an Alfv\'en wave will excite a compressive motion transverse
to the background magnetic field. The amplitude of this compressive motion
is second order in the amplitude $\varepsilon$ of the Alfv\'en wave itself.
This effect is somewhat analogous to the non-linear coupling between a 
fast mode and an Alfv\'en wave that is propagating through a 
non-relativistic magnetofluid with a gradient in density and in 
Alfv\'en speed (e.g. \cite{NK97}, \cite{TS02}).  In a spherical
spacetime, we show that a torsional wave originating at infinity is 
scattered into a spherical outgoing compressive wave.  

To this end, we expand the field and currents perturbations 
in powers of $\varepsilon \ll 1$:
\be
F_{\mu \nu} = F_{\mu \nu}^{(0)} + \delta F_{\mu \nu}^{(1)} + \delta F_{\mu \nu}^{(2)} + O({\varepsilon}^3);
\ee
\be
J^{\mu} = \delta J^{\mu(1)} + \delta J^{\mu(2)} + O({\varepsilon}^3).
\ee
Here, $\delta F^{(N)}_{\mu \nu} =  O({\varepsilon}^{N})$, etc.

When considering the effects of spacetime curvature, we also restrict
ourselves to axisymmetric perturbations.  In a static spacetime,
this means that $\delta r$ and $\delta\phi$ decouple to linear order. 
Such a linear coupling is re-established in a rotating spacetime.

\subsection{Torsional waves in a cylindrically symmetric spacetime}

\subsubsection{Static spacetime ($g_{t \phi} = 0$)}

Torsional deformations of the magnetic field are supported by
electric currents which,
to leading order in $\varepsilon$, propagate along the background magnetic
field.   We first derive the analog, in a curved cylindrical spacetime,
of the linear wave equation (\ref{alfwave}).  With this in hand, we
find solutions to  $J^{\mu}F_{\mu \nu} = 0$ to {\it second} order in 
$\varepsilon$.

The force-free equation implies $J^{r(1)} = 0$ to linear order
(Eq. [\ref{fflin}]).  Combining expression (\ref{jrexp}) with the
with the  Bianchi identity $\partial_t\delta F^{(1)}_{rz} = 
\partial_z\delta F^{(1)}_{rt}$ gives
\be\label{waveeq}
g^{tt}\partial^2_t\left(\delta F^{(1)}_{rz}\right) +
g^{zz}\partial^2_z\left(\delta F^{(1)}_{rz}\right) = 0.
\ee
The electric field $\delta F^{(1)}_{rt}$ satisfies the same
wave equation.  The solution is
\be
\delta F^{\pm}_{rz} = \delta F^{\pm}_{rz} (z \mp v_z t);\;\;\;\;\;\;
\delta F^{\pm}_{rt} = \delta F^{\pm}_{rt} (z \mp v_z t).
\ee
where
\be\label{vzval}
v_z \equiv \sqrt{-{g_{tt}\over g_{zz}}}.
\ee
The two fluctuating field components are related by
\be\label{anstz}
\delta F^{\pm}_{t r} =
\mp v_z \delta F^{\pm}_{z r}.
\ee
The net result is that the Alfv\'en wave  propagates along the
$z$-axis, to leading order in $\varepsilon$, with a phase speed 
equal to that of a null geodesic
that lies tangent to the background magnetic field.  

An Alfv\'en wave front propagating along the background magnetic field
will be bent as the result of the dependence
of $g_{tt}$ (and $v_z$) on radius.  This, in turn, forces a
radial oscillation of the magnetofluid that is second order in
$\varepsilon$.  Indeed, inspection of the force-free equation
\be
J^{t(1)} \delta F^{(1)}_{tr} + J^{z(1)} \delta F^{(1)}_{zr} 
+ J^{\phi (2)} F^{(0)}_{\phi r} = 0,
\ee
shows that the first two terms generally do not cancel.  
Using Eq. (\ref{anstz}) to eliminate $\delta F_{tr}$ in terms
of $\delta F_{zr}$, one obtains
\be\label{ggrad}
g^{rr}g^{zz} (\delta F_{zr}^{(1)})^2 \partial_r (\ln v_z)
+ 4 \pi J^{\phi(2)}F^{(0)}_{\phi r} = 0.
\ee
The radial fluctuation of the magnetofluid
supports the additional current perturbation,
\be\label{jtwo}
4\pi\sqrt{-g}J^{\phi(2)} =
- B_0\left\{{g_{rr}\over g_{tt}}\frac{{\partial}^2 \delta r}{\partial t^2} +
{g_{rr}\over g_{zz}}{\partial^2\delta r\over\partial z^2}
+ \frac{\partial}{\partial r}
\left[\frac{\sqrt{-g}}{g_{\phi \phi} g_{rr}} \frac{\partial}{\partial r}
\left(\delta r {g_{rr}g_{\phi\phi}\over\sqrt{-g}}\right)\right]\right\}.
\ee

Let us first consider the special case of a nearly flat cylindrical 
spacetime.  In such a background, $v_z = 1+2\lambda$ and Eq. (\ref{ggrad}) 
reduces to 
\be\label{nearlyFs}
\frac {1}{2 \pi} (\delta F^{(1)}_{zr})^2 \frac{d\lambda}{dr}  + 
J^{\phi(2)} F^{(0)}_{\phi r} = 0,
\ee
to lowest order in $\lambda$.  Re-expressing this equation in terms 
of the perturbations $\delta r$ and $\delta\phi$ gives
\be\label{nearly}
-{\partial^2\delta r\over\partial t^2} +
\frac{\partial}{\partial r} \left [ \frac{1}{r} \frac{\partial}{\partial r} 
(r \delta r) \right ] + {\partial^2\delta r\over\partial z^2} =
-2r^2 \frac{d\lambda}{dr}\,
\left(\frac{\partial \delta \phi}{\partial z}\right)^2.
\ee

The analogous expressions for the BTZ black string spacetime 
are not much more complicated. In this case, $v_z = Z^{1/2}
= (r^2/\ell^2-M)^{1/2}$, which vanishes at the horizon
$r_H = M^{1/2}\ell$.  The force-free equation reduces to
\be\label{btzggrad}
{1\over 2} (\delta F^{(1)}_{zr})^2
\partial_r Z
+ 4 \pi  J^{{\phi}(2)} F^{(0)}_{\phi r} = 0.
\ee
We then have 
\be
- \frac{1}{Z^2} \frac {{\partial}^2 \delta r}{\partial t^2} 
+ {1\over Z}\frac {{\partial}^2 \delta r}{\partial z^2}
+ \frac {\partial}{\partial r}
\left[\frac {Z}{r} \frac {\partial}{\partial r}
\left(\frac{r \delta r}{Z}\right)\right] =
-\frac{1}{Z} \frac{r^3}{l^2}
\left(\frac {\partial {\delta \phi}}{\partial z}\right)^2.
\ee

\subsubsection{Non-static spacetime $(g_{t \phi} \ne 0)$}\label{nonstatic}

When the spacetime rotates $(g_{t \phi} \ne 0$), 
one finds that the Alfv\'en wave ansatz 
(\ref{anstz}) fails to satisfy the force-free equation to linear order
in $\varepsilon$.  Physically, this is because the torsional and
radial modes of the magnetofluid are coupled through conservation
of angular momentum.  One must therefore introduce simultaneous
perturbations in $\delta r$, $\delta \phi$, and $\delta t$.  
In terms of these variables, the field perturbations become (to first order)
\be
\delta F^{(1)}_{r \phi} = 
\frac{\partial}{\partial r}(F^{(0)}_{r \phi} \delta r);\;\;\;\;\;\;
\delta F^{(1)}_{t \phi} = 
\frac{\partial (\delta r)}{\partial t}F^{(0)}_{r \phi};\;\;\;\;\;\;
\delta F^{(1)}_{z \phi} = 
\frac{\partial (\delta r)}{\partial z}F^{(0)}_{r \phi};
\ee
\be
\delta F^{(1)}_{rt} =
\frac{\partial (\delta\widetilde \phi)}{\partial t}F^{(0)}_{r\phi} -
{\partial\over\partial r}\left(\frac {g^{t \phi}}{g^{tt}} 
F^{(0)}_{r \phi} \delta r\right);\;\;\;\;\;\;
\delta F^{(1)}_{rz} = 
\frac{\partial (\delta\widetilde\phi)}{\partial z}F^{(0)}_{r \phi};\;\;\;\;\;\;
\delta F^{(1)}_{tz} = 
-{\partial\delta r\over\partial z}
\left(\frac {g^{t \phi}}{g^{tt}}\right) F^{(0)}_{\phi r}.
\ee
Here
\be
\delta\widetilde\phi \equiv \delta \phi 
- \left({g^{t \phi}\over g^{tt}}\right)\delta t 
= \delta \phi + \left({g_{t \phi}\over g_{\phi\phi}}\right)\delta t.
\ee
In this situation, the $t$-$\phi$ components of the metric 
are not diagonal:  $g^{tt} = g_{\phi\phi}/\det(t,\phi)$,
$g^{\phi\phi} = g_{tt}/\det(t,\phi)$, and $g^{t\phi} = 
-g_{t\phi}/\det(t,\phi)$, where $\det(t,\phi) \equiv
g_{tt}g_{\phi\phi}-g_{t\phi}^2$.  

The force-free equations are, again to linear order,
\be
J^{r(1)}F^{(0)}_{r\phi} = 0 = J^{r(1)}F^{(0)}_{rt},
\ee
or equivalently
\be\label{jreqzero}
J^{r(1)} = 0;
\ee
and
\be\label{ffwj}
J^{t(1)}F^{(0)}_{tr} + J^{\phi (1)}F^{(0)}_{\phi r} = 0.
\ee
Finally, the currents are given by
\ba
4\pi\sqrt{-g}J^{t(1)} &=& 
{\partial}_{r}
(\sqrt {-g} g^{tt} g^{rr} \delta F^{(1)}_{tr} +
\sqrt {-g} g^{t \phi} g^{rr} \delta F^{(1)}_{\phi r})\nn
&+& \partial_z(\sqrt{-g}g^{tt}g^{zz}\delta F^{(1)}_{t z}) +
\partial_z(\sqrt{-g}g^{t\phi}g^{zz}\delta F^{(1)}_{\phi z});
\ea
\be
4\pi\sqrt{-g}J^{r(1)} = 
{\partial}_{t}
(\sqrt {-g} g^{rr} g^{t\phi} \delta F^{(1)}_{r \phi} +
\sqrt {-g} g^{rr} g^{tt} \delta F^{(1)}_{rt}) +
{\partial}_{z}
(\sqrt {-g} g^{rr}g^{zz} \delta F^{(1)}_{rz});
\ee
and
\ba
4\pi\sqrt{-g}J^{\phi (1)} &=&
\partial_r(\sqrt {-g} g^{\phi \phi} g^{rr} \delta F^{(1)}_{\phi r} +
\sqrt {-g} g^{\phi t} g^{rr} \delta F^{(1)}_{tr}) -
\partial_t(\sqrt{-g}(g^{t\phi})^2\delta F^{(1)}_{\phi t})\nn
&+& \partial_t(\sqrt{-g}g^{\phi\phi}g^{tt}\delta F^{(1)}_{\phi t}) 
+ \partial_z(\sqrt{-g}g^{\phi\phi}g^{zz}\delta F^{(1)}_{\phi z})
+ \partial_z(\sqrt{-g}g^{\phi t}g^{zz}\delta F^{(1)}_{tz}).\nn
\ea

Substituting for the field perturbations, Eq. (\ref{jreqzero}) becomes
\be\label{fff}
g^{tt} \frac{{\partial}^2 (\delta \widetilde{\phi})}{\partial t^2}
+ g^{zz} \frac{{\partial}^2 (\delta \widetilde{\phi})}{\partial z^2}
= -g^{tt}\frac{\partial}{\partial r} 
\left( \frac{g_{t \phi}}{g_{\phi\phi}} \right)\,
\frac{\partial \delta r} {\partial t}.
\ee
One quickly obtains an integral of motion from this equation
when the gradient in the $z$-direction vanishes:
\be
{\partial\widetilde\phi\over\partial t} +
\delta r{\partial\over\partial r}\left({g_{t\phi}\over g_{\phi\phi}}\right)
= 0.
\ee
The torsional perturbation $\delta\phi$ is slaved to 
the compressive perturbation $\delta r$ through conservation of
angular momentum.
The manipulation of the force-free equation (\ref{ffwj})
is a bit more complicated; some details are given in Appendix B.  
The final result is
\ba\label{ffe}
\left(g^{tt}{\partial^2 \delta r\over \partial t^2} + 
g^{zz}{\partial^2 \delta r\over\partial z^2}\right)
F^{(0)}_{r \phi} &+& \frac{g_{\phi\phi}}{\sqrt{-g}}
\frac{\partial}{\partial r} 
\left[ \frac{\sqrt{-g}}{g_{\phi \phi} g_{rr}}
\frac{\partial}{\partial r}(\delta r F^{(0)}_{r \phi}) \right] \nn
&=& 
{g^{tt}g_{\phi\phi}\over g_{rr}}
\left[{\partial\widetilde\phi\over\partial t} +
\delta r{\partial\over\partial r}\left({g_{t\phi}\over g_{\phi\phi}}\right)
\right]
{\partial\over\partial r}\left({g_{t\phi}\over g_{\phi\phi}}\right) 
F^{(0)}_{r\phi}.
\ea

The wave operator on the left-hand side of Eqs. (\ref{fff}) and (\ref{ffe})
is now mixed with an additional term involving the other coordinate
perturbation.  This source term can be treated as a perturbation when 
the spacetime is slowly rotating ($g_{t\phi}^2 \ll |g_{tt}g_{\phi\phi}|$).
It is also worth noting a simplification which takes place
when the perturbations are uniform in $z$:  then
the right-hand side of Eq. (\ref{ffe}) vanishes and, effectively,
the fast mode equation involves only a single coordinate perturbation.
That is, the time-evolution of the angular perturbation $\delta\phi$ 
is fixed by conservation of angular momentum, and does not react back 
on the equation of motion for $\delta r$.  More generally this is 
not the case, and the dynamics involves two independent fields.

\subsection{Scattering of a cylindrical Alfv\'en wave in a spherical
gravitational field}

A uniform magnetofluid is perturbed by a spherical gravitational
field (Section \ref{magsphere}).  We now consider the perturbation
to a torsional Alfv\'en wave propagating along this background
magnetic field.  We focus on a weak, spherical gravitational field,
and allow for the possibility that the gravitating mass is extended in
radius.  Then the metric departs from the Schwarzschild metric;  
we write it as
\be
-g_{tt} = 1 + \Delta_{tt};\;\;\;\;\;\; g_{rr} = 1 + \Delta_{rr}.
\ee
(In this section, as in Section \ref{magsphere}, $r$ is the spherical radius.)
In the weak field limit ($|\Delta_{tt}|$, $|\Delta_{rr}| \ll 1$) 
the radial coordinate
of the magnetofluid is perturbed by an amount $\delta r/r = O(\Delta)$.

The static perturbation $\delta r$ is a solution to the
linearized version of Eq. (\ref{Rfeq}),
\be\label{deltareq}
{d^2(r\delta r)\over dr^2} - 2{(r\delta r)\over r^2} = 
{1\over 2}(\Delta_{tt} + \Delta_{rr}) - {1\over 2}{d\over dr}
\left[r\left(\Delta_{tt} - \Delta_{rr}\right)\right].
\ee
Notice that the right-hand side of this equation vanishes in
the Schwarzschild metric, $\Delta_{rr} = -\Delta_{tt} = 2GM/r$.
When the enclosed gravitating mass $M$ is itself a function of 
radius, Eq. (\ref{deltareq}) becomes
\be\label{deltareqb}
{d^2(r\delta r)\over dr^2} - 2{(r\delta r)\over r^2} = 
2G{dM\over dr}.
\ee

We expand the background magnetic field $F^{(0)}$,
the Alfv\'en wave $\delta F^{(1)}$, and its current
$\delta J^{(1)}$ in powers of $\Delta$, 
\be
F^{(0)} = F^{(0)}_{0-\Delta} + F^{(0)}_{1-\Delta}
\ee
and
\be
\delta F^{(1)} = \delta F^{(1)}_{0-\Delta} + \delta F^{(1)}_{1-\Delta};
\;\;\;\;\;\;
\delta J^{(1)} = \delta J^{(1)}_{0-\Delta} + \delta J^{(1)}_{1-\Delta},
\ee
so that $F^{(0)}_{1-\Delta}$, $\delta F^{(1)}_{1-\Delta}$ and
$\delta J^{(1)}_{1-\Delta}$ are
first order in $\delta r$.  For example, expanding
${\cal R} = r_0^2 \simeq r^2 + 2r\delta r$
in Eq. (\ref{fpert}) for the background field, we have
\be\label{F0}
\left[F^{(0)}_{0-\Delta}\right]_{r\phi} = B_0 \,r\,\sin^2\theta;
\;\;\;\;\;\;
\left[F^{(0)}_{0-\Delta}\right]_{\theta\phi} = 
B_0 \,r^2\,\sin\theta\cos\theta,
\ee
and
\be\label{deltaF1}
\left[F^{(0)}_{1-\Delta}\right]_{r\phi} = 
B_0 \,{d(r\delta r)\over dr}\,\sin^2\theta;
\;\;\;\;\;\;
\left[F^{(0)}_{1-\Delta}\right]_{\theta\phi} = 
2B_0 \,r\delta r\,\sin\theta\cos\theta.
\ee

In the absence of gravity, the torsional wave propagates along
the magnetic field. The associated charge and current densities 
\ba\label{jvsrcyl}
\delta\rho &=& \delta J_0(z-t)\,
\exp\left[-{1\over 2}\left({r\sin\theta\over R_0}\right)^2\right];\nn
\delta{\bf J} &=& \delta J_0(z-t)\,
\exp\left[-{1\over 2}\left({r\sin\theta\over R_0}\right)^2\right]\hat z
\ea
are localized within a cylindrical radius
$r\sin\theta \sim R_0$.  

We will focus on the case where the
wave is strongly sheared, i.e., where its frequency is small compared
with $R_0^{-1}$:
\be
R_0\left|{\delta J_0'(z-t)\over\delta J_0(z-t)}\right| \ll 1.
\ee
To lowest order in the gravitational potential, the components
of the fluctuating current and field are
\be\label{deltaJ1}
\left[\delta J^{(1)}_{0-\Delta}\right]^t = \delta J;\;\;\;\;\;\;
\left[\delta J^{(1)}_{0-\Delta}\right]^r = \cos\theta\,\delta J;\;\;\;\;\;\;
\left[\delta J^{(1)}_{0-\Delta}\right]^\theta = 
-{1\over r}\sin\theta\,\delta J;
\ee
and
\ba
\left[\delta F^{(1)}_{0-\Delta}\right]_{\theta r}
&=& -{4\pi \delta J_0\,R_0^2\over\sin\theta}\,
\left\{1- \exp\left[-{1\over 2}\left({r\sin\theta\over R_0}\right)^2\right]
\right\};\nn
\left[\delta F^{(1)}_{0-\Delta}\right]_{t r} &=&
{\sin\theta\over r}\left[\delta F^{(1)}_{0-\Delta}\right]_{\theta r};
\;\;\;\;\;\;
\left[\delta F^{(1)}_{0-\Delta}\right]_{t\theta} =
\cos\theta\left[\delta F^{(1)}_{0-\Delta}\right]_{\theta r}.
\ea

The static gravitational perturbation to the background magnetic field
is known implicitly through the solution to Eq. (\ref{deltareq}).
We next solve for the perturbations $\delta F^{(1)}_{1-\Delta}$
and $\delta J^{(1)}_{1-\Delta}$.  The current perturbation is
determined from the linear component of the force-free equation.
Making use of $\left[\delta J^{(1)}_{0-\Delta}\right]^\mu
\left[F^{(0)}_{0-\Delta}\right]_{\mu\nu} = 0$ and neglecting
the terms that are second-order in $\Delta$, we have
\be\label{fflinear}
\left[\delta J^{(1)}_{0-\Delta}\right]^\mu 
\left[F^{(0)}_{1-\Delta}\right]_{\mu\nu}
+ \left[\delta J^{(1)}_{1-\Delta}\right]^\mu
\left[F^{(0)}_{0-\Delta}\right]_{\mu\nu}
= 0.
\ee
The force-free equation (\ref{fflinear}) may be re-expressed by
substituting Eqs. (\ref{F0}), (\ref{deltaF1}) and (\ref{deltaJ1}),
\be\label{fflinearb}
\left[\delta J^{(1)}_{1-\Delta}\right]^r \sin\theta +
\left[\delta J^{(1)}_{1-\Delta}\right]^\theta r\cos\theta 
= -\delta J(r\sin\theta,z,t)\,\left[ r{d(\delta r/r)\over dr}\right]\,
\sin\theta\cos\theta.
\ee
In the low frequency regime, the equation of current conservation
simplifies to
\be\label{cons}
{1\over r^2}{\partial\over\partial r}\left\{r^2
\left[\delta J^{(1)}_{1-\Delta}\right]^r\right\} +
{1\over\sin\theta}{\partial\over\partial\theta}\left\{
\sin\theta\left[\delta J^{(1)}_{1-\Delta}\right]^\theta\right\} = 0.
\ee
For the particular choice (\ref{jvsrcyl}) of incident current density,
the solution to the combined equations (\ref{fflinearb}) and (\ref{cons}) is
\be
\left[\delta J^{(1)}_{1-\Delta}\right]^r = \left[2{\delta r\over r}
\cos\theta - {r\delta r\over R_0^2}\cos\theta\sin^2\theta\right]\delta J
\ee
and
\be
\left[\delta J^{(1)}_{1-\Delta}\right]^\theta = 
\left[-{1\over r^2}{d(r\delta r)\over dr}\sin\theta
+{\delta r\over R_0^2}\sin^3\theta\right]\delta J.
\ee
Notice that the current $\delta J^{(1)}_{1-\Delta}$ 
receives a contribution from the metric perturbation as well
as from $\delta F^{(1)}_{1-\Delta}$, e.g.
\be
4\pi\left[\delta J^{(1)}_{1-\Delta}\right]^r = 
{1\over r^2\sin\theta}{\partial\over\partial\theta}\left\{
\sin\theta (g^{rr}-1)\left[\delta F^{(1)}_{0-\Delta}\right]_{r\theta}\right\}
+ {1\over r^2\sin\theta}{\partial\over\partial\theta}\left\{
\sin\theta \left[\delta F^{(1)}_{1-\Delta}\right]_{r\theta}\right\}.
\ee
Here, the metric perturbation is $g^{rr}-1 = -2GM/r$ to lowest order. 
Thus the
current components $[\delta J^{(1)}_{1-\Delta}]^{r,\theta}$ are sourced by
the fluctuating field 
\be
\left[\delta F^{(1)}_{1-\Delta}\right]_{\theta r}
= -\left[4\pi (r\delta r)\sin\theta\right]\,\delta J(r\sin\theta,z,t)
+ {2GM\over r}\left[\delta F^{(1)}_{0-\Delta}\right]_{\theta r}.
\ee
The gravitational perturbation of the charge density,
and the electric field which it sources, can be obtained by combining
the Bianchi identity
\be\label{bianchi}
\partial_\theta\left[\delta F^{(1)}_{1-\Delta}\right]_{tr} -
\partial_r\left[\delta F^{(1)}_{1-\Delta}\right]_{t\theta}
= \partial_t\left[\delta F^{(1)}_{1-\Delta}\right]_{\theta r}
\simeq 0,
\ee
with the MHD condition
\ba
&&\left\{\left[\delta F^{(1)}_{0-\Delta}\right]_{t\theta} + 
       \left[\delta F^{(1)}_{1-\Delta}\right]_{t\theta}\right\}
\left\{\left[F^{(0)}_{0-\Delta}\right]_{\phi r} + 
       \left[F^{(0)}_{1-\Delta}\right]_{\phi r}\right\}\nn
&-&
\left\{\left[\delta F^{(1)}_{0-\Delta}\right]_{tr} + 
       \left[\delta F^{(1)}_{1-\Delta}\right]_{tr}\right\}
\left\{\left[F^{(0)}_{0-\Delta}\right]_{\phi\theta} + 
       \left[F^{(0)}_{1-\Delta}\right]_{\phi\theta}\right\} = 0.
\ea
At leading order in the gravitational potential,
this last equation becomes
\be
\left[\delta F^{(1)}_{1-\Delta}\right]_{t\theta} = 
r\cot\theta\,\left[\delta F^{(1)}_{1-\Delta}\right]_{tr}
- \cos\theta {d(r^3\delta r)\over d\ln r}\,
\left[\delta F^{(1)}_{0-\Delta}\right]_{\theta r}.
\ee
Substituting this expression into Eq. (\ref{bianchi}), one
can solve for the perturbation to the electric field,
\be
\left[\delta F^{(1)}_{1-\Delta}\right]_{tr} = 
-4\pi\delta J_0\left[
R_0^2{d(\delta r/r)\over dr}
\left(1-e^{-R^2/2R_0^2}\right) + \sin^2\theta\,\delta r
e^{-R^2/2R_0^2}\right],
\ee
and
\be
\left[\delta F^{(1)}_{1-\Delta}\right]_{t\theta} = 
-[4\pi (r\delta r)\,\sin\theta\cos\theta]\delta J_0 e^{-R^2/2R_0^2}.
\ee
Here $R = r\sin\theta$. The charge density perturbation is then
\ba
4\pi\left[\delta J^{(1)}_{1-\Delta}\right]^t
= -{1\over r^2}{\partial\over\partial r}\left\{
r^2\left[\delta F^{(1)}_{1-\Delta}\right]_{tr}\right\}
&-&{1\over r^2\sin\theta}{\partial\over\partial\theta}\left\{
\sin\theta\left[\delta F^{(1)}_{1-\Delta}\right]_{t\theta}\right\}\nn
&+&{1\over r^2\sin\theta}{\partial\over\partial\theta}\left\{
\sin\theta\left(g^{tt}+1\right)\,
\left[\delta F^{(1)}_{0-\Delta}\right]_{t\theta}\right\}.
\ea
Here $g^{tt}+1 = -2GM/r$ to lowest order.

Our principal goal here is to work out the second-order current
$[\delta J^{(2)}_{1-\Delta}]^\phi$.  This may be obtained from
\be\label{ffquad}
\left[\delta J^{(1)}_{0-\Delta}\right]^\mu 
\left[\delta F^{(1)}_{1-\Delta}\right]_{\mu\nu}
+ \left[\delta J^{(1)}_{1-\Delta}\right]^\mu
\left[\delta F^{(1)}_{0-\Delta}\right]_{\mu\nu}
= -\left[\delta J^{(2)}_{1-\Delta}\right]^\phi
\left[F^{(0)}_{0-\Delta}\right]_{\phi\nu},
\ee
by substituting Eqs. (110), (111), (116), (118), (122) and (124).
The resulting expression is fairly complicated, but it simplifies
dramatically when the gravitating mass is concentrated at a radius
much smaller than the transverse scale $R_0$ of the incident Alfv\'en
wave.  In this regime, one has $\delta r/r \sim G(dM/dr) \ll GM/r$,
and the dominant field and current perturbations are
\ba
\left[\delta F^{(1)}_{1-\Delta}\right]_{\theta r}
&\simeq& {2GM\over r}\left[\delta F^{(1)}_{0-\Delta}\right]_{\theta r}\nn
&=& \left({2GM\over r}\right)\,{4\pi R_0^2\over\sin\theta}\,
\left[1- e^{-R^2/2R_0^2}\right]\,\delta J_0(r\cos\theta-t),
\ea
and
\be
\left[\delta J^{(1)}_{1-\Delta}\right]^t
\simeq {2GM\over r}\left[\cos^2\theta\,e^{-R^2/2R_0^2} -
{R_0^2\over r^2}\left(1-e^{-R^2/2R_0^2}\right)\right]\,
\delta J_0(r\cos\theta-t).
\ee
Combining the terms on the left-hand side of Eq. (\ref{ffquad}), 
and substituting the background field (\ref{fminksph}), we find
\be
\left[\delta J^{(2)}_{1-\Delta}\right]^\phi
= {8\pi(\delta J_0)^2\over B_0\sin^2\theta}\,
\left({GM\over r}\right)\,
\left({R_0\over r}\right)^4\,\left(1-e^{-R^2/2R_0^2}\right)^2
\left[1 + {R^2/R_0^2\over e^{R^2/2R_0^2}-1}\right].
\ee
This expression simplifies further
at small cylindrical radius ($R \ll R_0$),
\be
\left[\delta J^{(2)}_{1-\Delta}\right]^\phi \simeq 
6\pi{|\delta J_0|^2\over B_0}\left({GM\over r}\right)\sin^2\theta\,
e^{2i\omega(z-t)},
\ee
when the incident Alfv\'en wave is a pure harmonic.
The second-order current is the source for an outgoing compressive wave
\be\label{spherewave}
{\partial^2(r\delta r_I)\over\partial t^2}
- {1\over r^2\sin\theta}{\partial\over\partial\theta}
\left(\sin\theta{\partial(r\delta r_I)\over\partial\theta}\right)
- {\partial^2(r\delta r_I)\over\partial r^2} =
{4\pi r^2\over B_0}\left[\delta J^{(2)}_{1-\Delta}\right]^\phi,
\ee
of amplitude $\delta r_I$.
As in the case of a torsional wave propagating in a cylindrical
spacetime, the compressive mode is second harmonic and second-order 
in the wave amplitude.  

The Greens function solution to Eq. 
(\ref{spherewave}) is dominated by the fluid at radius
$r \sim R_0$.  As a result, the shape of the asymptotic
fast wave is sensitive to the distribution of current within the
incoming Alfv\'en mode.  We will not examine the pattern of
the outgoing compressive wave in full detail here.

\section{Collisions of Axisymmetric Waves in a Cylindrical Spacetime}

In this section, we study the collision of two torsional Alfv\'en waves 
in a uniform magnetofluid in a static cylindrical spacetime 
($g_{t\phi} = 0$).  
The effects of spacetime curvature are subtle enough that
we specialize to axisymmetric modes, supported only by a perturbation
$\delta\phi$.  The result easily generalizes
to the case of flat space, and in the next section we examine
more general types of wave interactions in a Minkowski background.

The two Alfv\'en waves, propagating oppositely along the magnetic field,
are labeled $+$ and $-$, and the component of the current generated
by their interaction is labelled $I$.  
The colliding waves are each assumed to be supported within a 
cylindrical shell,  between radii $R_{\rm min}$ and $R_{\rm max}$.  
We start from the force-free equation
\be\label{crossterms}
(J^t_+ + J^t_-)(\delta F_{tr}^+ + \delta F_{tr}^-)
 + (J^z_+ + J^z_-)(\delta F_{zr}^+ + \delta F_{zr}^-)
+ J^\phi_I F_{\phi r}^{(0)}  = 0.
\ee
The background magnetic field is defined by $J^{\phi(0)} = 0$ [Eq.
(\ref{bback})].  The fluctuating fields are 
\ba\label{dftr}
\delta {F_{zr}}^{\pm} &=&
\frac{\partial {\phi}_{\pm}}{\partial z} F^{(0)}_{\phi r};\nn
\delta {F_{tr}}^{\pm} &=&
\frac{\partial {\phi}_{\pm}}{\partial t} F^{(0)}_{\phi r}
= \mp v_z \delta {F_{zr}}^{\pm},
\ea
where $v_z$ is given by Eq. (\ref{vzval}).  The associated currents are
\be
J^t_{\pm} = {1\over 4\pi \sqrt{-g}}{\partial\over\partial r}
\left(\sqrt{-g} g^{tt}g^{rr}\delta F^{\pm}_{tr}\right)
= \pm {1\over 4\pi \sqrt{-g}}{\partial\over\partial r}
\left(\sqrt{-g} {g^{zz}g^{rr}\over v_z}\delta F^{\pm}_{zr}\right)
\ee
and
\be
J^z_{\pm} = {1\over 4\pi \sqrt{-g}}{\partial\over\partial r}
\left(\sqrt{-g} g^{zz}g^{rr}\delta F^{\pm}_{zr}\right).
\ee
Combining these expressions, the force-free equation becomes
\ba\label{alfcol0}
-{4\pi\over B_0}\sqrt{-g} J^\phi_I 
&=& {g_{rr}\over g_{tt}}\frac{{\partial}^2 \delta r}{\partial t^2} +
{g_{rr}\over g_{zz}}{\partial^2\delta r\over\partial z^2}
+ \frac{\partial}{\partial r}
\left[\frac{\sqrt{-g}}{g_{\phi \phi} g_{rr}} \frac{\partial}{\partial r}
\left(\delta r {g_{rr}g_{\phi\phi}\over\sqrt{-g}}\right)\right]\nn
&=& -2\left({g_{zz}\over g_{\phi\phi}}\right)
{\partial\over\partial r}\left[\left({g_{\phi\phi}\over g_{zz}}\right)^2
{\partial\phi_+\over\partial z}{\partial\phi_-\over\partial z}\right] +\nn
&&\left({g_{\phi\phi}\over g_{zz}}\right)\,\left[
2{\partial\phi_+\over\partial z}{\partial\phi_-\over\partial z}
-\left({\partial\phi_+\over\partial z}\right)^2 
-\left({\partial\phi_-\over\partial z}\right)^2\right]
\partial_r(\ln v_z).
\ea
In contrast with the case of a single (unidirectional) Alfv\'en
wave, a source term appears 
even when the wave speed has a vanishing gradient ($\partial_r v_z = 0$).
For that reason we must, in general, retain the curvature corrections
in the wave operator for $\delta r$.

\subsection{Collision in a nearly flat, cylindrically symmetric spacetime}
\label{acol}

We now specialize to a nearly flat cylindrical spacetime with 
Newtonian potential $\lambda$ and metric (\ref{weak}), and
focus on the source terms in Eq. (\ref{alfcol0}) that
are proportional to $(\partial_z\phi_+)(\partial_z\phi_-)$.
The torsional wave speed is $v_z = 1+2\lambda$.  Comparing with
flat space, the new effects we are interested in
are already present in the case of
a very long wavelength twist in the magnetic field,
\be
\left|{\partial\phi_\pm\over\partial z}\right| R_{\rm max} \ll 1,
\ee
so we neglect the derivatives in $t$ and $z$. The perturbation to $B^z$ is
\be\label{dbz}
{\delta B^z\over B_0} = {\delta F^I_{\phi r}\over F_{\phi r}^{(0)}}
= {1\over r}{\partial(r\delta r)\over\partial r}
-2{d\lambda\over dr}\delta r,
\ee
and the equation of motion (\ref{alfcol0}) simplifies to
\be\label{alfcol}
{\partial\over\partial r}\left({\delta B^z\over B_0}\right)
= -{2\over r^2} {\partial\over\partial r}
\left ( r^4 \frac {\partial \phi_{+}}{\partial z}
\frac {\partial \phi_{-}}{\partial z} \right ) +
4 r^2{d\lambda\over dr}
{\partial\phi_+\over\partial z}{\partial\phi_-\over\partial z}.
\ee

Our first goal is to calculate the amplitude $
\delta R_{\rm max}(t) \equiv \delta r(R_{\rm max},t)$ of
the hydrodynamic fluctuation at the outer boundary of the annulus
supporting the Alfv\'en waves.  The amplitude of the fast mode
at large radius can then easily be obtained, to leading order
in $\lambda$, from the equation of motion (\ref{rmink}) in Minkowski space.  
Taking the two waveforms to be \footnote{Generally the wavenumbers
$k_+$ and $k_-$ are not equal; but the translational invariance of 
the background magnetofluid in the $z$-direction allows them
to be made equal through an appropriate Lorentz boost.}
\be
\delta \phi_\pm(t,r,z) = \delta\phi_{0\pm}\sin\Bigl[k(z\mp t)\Bigr]
\,f_{\perp\pm}(r),
\ee
one sees that the interaction term
\be
{\partial\phi_+\over\partial z}{\partial\phi_-\over\partial z}
= {1\over 2}k^2\delta\phi_{0+}\delta\phi_{0-}\Bigl[
\cos(2kt) + \cos(2kz)\Bigr]f_{\perp+}(r)f_{\perp-}(r)
\ee
is the sum of a uniform oscillation of the cylinder, and a
static deformation that excites no wave motion outside the cylinder. 
Thus the calculation of the fast mode amplitude reduces to 
the problem of calculating the amplitude $\delta R_{\rm max}$
of the oscillation at the outer boundary of the cylinder.  

We start by integrating Eq. (\ref{dbz}) outward from 
$\delta R = 0$ at $r = 0$ to obtain
\be
R_{\rm min} \delta R_{\rm min} = 
\left({1\over 2}R_{\rm min}^2 + \int_0^{R_{\rm min}}
{d\lambda\over dr} r^2 dr\right){\delta B^z\over B_0}\biggr|_{R_{\rm min}},
\ee
to first order in $\lambda$.  
The expansion of the magnetofluid is nearly incompressible in the near zone
outside the cylinder ($kr \ll 1$), so the relevant inner and outer boundary 
conditions are
\be
{\delta B^z\over B_0} \sim (kR_{\rm max})^2{\delta R_{\rm max}\over
R_{\rm max}} \simeq 0,
\ee
and 
\be
{\partial\phi_\pm\over\partial z}\biggr|_{R_{\rm max}} = 
{\partial\phi_\pm\over\partial z}\biggr|_{R_{\rm min}} = 0.
\ee
The field perturbation is nearly uniform
inside radius $R_{\rm min}$; we will only need the leading term
\be
{\delta B^z\over B_0}\biggr|_{R_{\rm min}} =
0 + 2\int_{R_{\rm min}}^{R_{\rm max}} {1\over r^2}{\partial\over\partial r}
\left(r^4 {\partial\phi_+\over\partial z}
{\partial\phi_-\over\partial z}\right)dr = 4\Phi_{_{+-}}(R_{\rm max}),
\ee
where 
\be
\Phi_{_{+-}}(r) \equiv \int_{R_{\rm min}}^r r{\partial\phi_+\over\partial z}
{\partial\phi_-\over\partial z}dr.
\ee
The radial perturbation within the annulus $R_{\rm min} < r < R_{\rm max}$
can be found iteratively from
\ba
r\delta r &-& R_{\rm min}\delta R_{\rm min} = 
\int_{R_{\rm min}}^r r'\left[{\delta B^z\over B_0} + 2{d\lambda\over dr'}
\delta r(r')\right]dr'\nn
&=& - {1\over 2}R_{\rm min}^2
{\delta B^z\over B_0}\biggr|_{R_{\rm min}} + 
{1\over 2}r^2{\delta B^z\over B_0} - \int_{R_{\rm min}}^r
{1\over 2}r'^2{\partial\over\partial r'}\left({\delta B^z\over B_0}\right)dr'
+2\int_{R_{\rm min}}^r r'{d\lambda\over dr'}\delta r(r')dr',\nn
\ea
after substituting Eq. (\ref{alfcol}) in the third term on the right-hand
side. 
Evaluating this expression at $r = R_{\rm max}$ we find
\be
R_{\rm max}\delta R_{\rm max} = 
-2\int_{R_{\rm min}}^{R_{\rm max}} r^2{d\lambda\over dr}
\left(2\Phi_{_{+-}}(r)+r{d\Phi_{_{+-}}\over dr}\right)dr
+4 \Phi_{_{+-}}(R_{\rm max})\int_0^{R_{\rm max}}r^2{d\lambda\over dr}dr.
\ee

The first remark to make about this expression is that
$\delta R_{\rm max}$ vanishes in flat space ($d\lambda/dr =$ constant).
The same turns out to be true in
the special case of a cylindrical line mass ($d\lambda/dr = K/r$),
but not for a more general cylindrical mass distribution that is
extended in radius.  

We now obtain the fast mode amplitude in the wave zone $kr \gg 1$.
The cylindrical wave equation for $\delta B^z$ has
the outgoing wave solution 
\be
\frac{\delta B^z}{B_0} = A e^{-i k t} {H_0}^{(1)}(kr) \sim
A \sqrt{2\over\pi kr}e^{ik(r-t)-i\pi/4}\;\;\;\;\;\;(kr \gg 1)
\ee
in Minkowski space.  
The Hankel function scales as $H_0^{(1)}(kr) \sim (2i/\pi)\ln(kr)$ at
small radius.
To obtain the normalization factor, we note that in the near
zone ($k r \ll 1$) where the magnetofluid is nearly incompressible,
the rate of transport of magnetic flux is nearly constant in radius,
\be \label{flux}
{\partial\Phi\over\partial t}=2 \pi r {\partial(\delta r)\over\partial t}
B^z \simeq {\rm constant} \;\;\;\;\;\; (kr \ll 1).
\ee
This implies
\be\label{vr} 
{\partial \delta r\over\partial t} = 
\left [ 
\frac {\partial (\delta r)}{\partial t}
\right]_{R_{\rm max}}
\left (
\frac{r}{R_{\rm max}} 
\right )^{-1}.
\ee
Substituting this expression into $\vec{E} = - 
\partial_t(\delta r) B_0(\hat{r} \times \hat{z})$,
and thence into
\be
-\frac {\partial \vec{E}}{\partial t} = \vec{\nabla} \times \vec{B},
\ee
and integrating over radius, gives
\be
\frac {\delta B^z}{B_0} - \frac{\delta B^z(R_{\rm max})}{B_0} =
\left[ \frac {{\partial}^2 (\delta r)}{\partial t^2}
\right]_{R_{\rm max}} 
\ln\left({r\over R_{\rm max}}\right) R_{\rm max},
\ee
and 
\be
Ae^{-ikt} = {i\pi\over 2}\Bigl(\omega^2 R_{\rm max}\Bigr)\,
\delta r(R_{\rm max},t).
\ee
The general solution is
\be
\frac{\delta B^z(r,t)}{B_0} = - R_{\rm max}\int_{-\infty}^{t - r}
\frac{\partial^2(\delta r)}{\partial t'^2}
\frac{dt'}{\sqrt{(t - t')^2 - r^2}}.
\ee

\subsection{Collision in a static black string spacetime}

As another illustrative example, we consider the collision between
two torsional waves in the spacetime of a static black string 
[Eq. (\ref{dsbtzj}) with $J = 0$].  The background magnetic field 
$F^{(0)}_{r\phi} = B_0 r / (r^2/\ell^2-M)$ is aligned with the axis
of the string.  We assume, as before, that the colliding waves are supported
only within a cylindrical annulus $R_{\rm min} < r < R_{\rm max}$,
and take $R_{\rm min}$, $R_{\rm max} \sim \ell$.

A radial disturbance of frequency $\omega \ll \ell^{-1}$ accumulates a phase
\be
\phi \sim \omega \int \left(-{g_{rr}\over g_{tt}}\right)^{1/2}dr
= \omega \int {dr\over r^2/\ell^2-M}.
\ee
This integral converges at large radius but diverges near the horizon
$r_H = M^{1/2}\ell$.  When $M \ll (\omega \ell)^2$,  there is a
well defined wave zone which is localized at a {\it small} radius,
$r \lsim \omega \ell^2$.
The effect of a small finite angular momentum $J$ is to impose a reflecting
barrier for the magnetosonic wave at a radius $r \sim (J^2/4\omega)^{1/3}$,
and therefore to form a resonant cavity.  
We will not consider the case of finite angular momentum here.

The medium outside radius $R_{\rm max}$ will, at the same time, 
respond to a low-frequency disturbance with a uniform compression or
rarefaction, $\delta F_{\phi r} \sim F_{\phi r}^{(0)}$.  
Thus $\delta r \sim r$ at a large radius ($r \gg \ell$).  The amplitude of 
the radial perturbation $\delta R_{\rm max}$ can be obtained by
noting that the magnetic flux 
\be
\Phi
= \int 2\pi dr F_{r\phi} = \int dr {2\pi rB_0\over r^2/\ell^2-M}
\ee
diverges at large radius.  At the same time, the background field energy
\be
{dE_B\over dz} = 
\int 2\pi dr \sqrt{g_{rr}g_{\phi\phi}} \left[-g^{tt}g^{rr} g^{\phi\phi} 
{F_{r\phi}^2\over 8\pi}\right]
= \int {B_0^2\over 4} {r dr\over (r^2/\ell^2-M)^{5/2}}
\ee
is finite.  As a result, an oscillation $\delta R_{\rm max}$ of
the outer boundary of the annulus is accompanied by a small
field perturbation, $\delta B_z(R_{\rm max}) \simeq 0$.

The nature of the wave solutions at small radius is most easily demonstrated
by taking the limit $M\rightarrow 0$, $Z(r) \rightarrow r^2/\ell^2$.
Then the wave equation (\ref{waveBTZ}), which applies to
the case of a uniform radial oscillation
$\delta r(R_{\rm max},z,t) \propto  e^{-i\omega t}$, simplifies
considerably.   It transforms under the
change of variable $r_* = -\ell^2/r$ to
\be
- \frac{{\partial}^2}{\partial t^2} 
\left ( \frac {\delta B^z}{B^z} \right ) + 
r_* \frac{\partial}{\partial r_*} 
\left[\frac{1}{r_*} \frac{\partial}{\partial r_*} 
\left(\frac{\delta B^z}{B^z}\right)\right] = 0.
\ee
A further change of variable 
$\delta B^z/B^z = h(r_*) (\omega r_*) e^{-i \omega t}$ yields
the Bessel equation
\be
r_*^2 \frac{{\partial}^2 h}{\partial r_*^2} + 
r_* \frac{\partial h}{\partial r_*} + (\omega^2 r_*^2 - 1) h = 0,
\ee
which has the wave solution
\be
\frac {\delta B^z}{B^z} = (\omega r_*)
H_1^{(2)}(\omega r_*)\,Ae^{-i\omega t}  \sim 
A (\omega r_*)\sqrt{ \frac{2}{\pi (\omega r_*)}}  
e^{-i\omega(r_* + t) - 3\pi/4}
\ee
propagating to small $r$ (large negative $r_*$).
It will be noted that the amplitude of the wave {\it diverges}
in the wave zone, because the radial phase velocity
\be
{\omega\over k} = \sqrt{-{g_{tt}\over g_{rr}}} \sim 
\left({\ell\over r_*}\right)^2
\ee
asymptotes to zero.  

To relate the coefficient $A$ to the 
amplitude of the oscillation at radius $R_{\rm min}$, we note
that the rate of transport of magnetic flux
\be
2\pi F_{r\phi} {\partial(\delta r)\over\partial t} =
{2\pi B_0\ell^2\over r} {\partial(\delta r)\over\partial t}
\ee
is constant in the near zone ($|\omega r_*| \ll 1$).
Hence $\partial_r(\delta r) \simeq \delta r/r$.  From 
Eq. (73) we have,
\be
{\partial^2\delta r_*\over \partial t^2} = {\partial\over\partial r_*}
\left( {\delta B^z\over B^z}\right).
\ee
Combining this with the low frequency
expansion $(\omega r_*)H_1^{(2)}(\omega r_*) \sim 2i/\pi
-(i/\pi)(\omega r_*)^2 \ln(\omega r_*/2)$ gives
\be
Ae^{-i\omega t} \simeq {-i\pi/2\over \ln(\omega R^*_{\rm min}/2)}
{\delta R_{\rm min}^*(t)\over R_{\rm min}^*}.
\ee

The last step in obtaining the amplitude of the fast mode
that emerges from the wave collision, is to express $\delta R_{\rm min}$
in terms of the amplitudes of the colliding Alfv\'en modes.
The terms involving $(\partial_z\phi_+)(\partial_z\phi_-)$
in Eq. (\ref{alfcol0}) can be combined to give
\be\label{derivcollision}
\frac {\partial}{\partial r} \left(\frac {\delta B^z}{B^z}\right) =
\frac {\partial}{\partial r} \left[ r \frac {\partial}{\partial r}
\left( {\delta r\over r} \right) \right] =
-{2\over r}{\partial\over\partial r}\left ( r^3
\frac {\partial \phi_+}{\partial z}
\frac {\partial \phi_-}{\partial z} \right).
\ee
Integrating with respect to radius, this becomes
\ba\label{inptgenrldrBTZ}
{\delta R_{\rm min}\over R_{\rm min}} - 
{\delta R_{\rm max}\over R_{\rm max}}
&=& -\int_{R_{\rm min}}^{R_{\rm max}} dr
\left\{{\partial\over\partial r}
\left[\ln\left({r\over R_{\rm min}}\right)\,{\delta B^z\over B^z}\right]
-\ln\left({r\over R_{\rm min}}\right)\,
{\partial\over\partial r}\left({\delta B^z\over B^z}\right)\right\}\nn
&=& 2\int_{R_{\rm min}}^{R_{\rm max}} dr
r\left[1-\ln\left({r\over R_{\rm min}}\right)\right]
{\partial\phi_+\over\partial z}{\partial\phi_-\over\partial z},
\ea
since $\partial_z\phi_\pm(R_{\rm min}) = 
\partial_z\phi_\pm(R_{\rm max}) = 0$ and $\delta B_z(R_{\rm max}) = 0$.
The field perturbation at the inner boundary is obtained by
substituting $R_{\rm min} \rightarrow R_{\rm max}$ in the logarithms
in the first line of Eq. (\ref{inptgenrldrBTZ}).  
After subtracting the two equations, one obtains
$$
{\delta B^z\over B^z}\biggr|_{R_{\rm min}} = 
2\int_{R_{\rm min}}^{R_{\rm max}} r {\partial \phi_+\over\partial z}
{\partial\phi_-\over\partial z} dr.\eqno(166b)
$$

\section{Collisions of Non-Axisymmetric Waves in Minkowski Space}

We will now broaden our discussion of mode collisions to non-axisymmetric
modes, and consider both fast and Alfv\'en modes.  Three-mode interactions
in the force-free limit were previously considered in \cite{TB98},
using the magnetic Lagrangian formalism developed there.  We return
to this subject and clarify the properties of i)
the three-mode coupling between two Alfv\'en waves;
ii) the interaction between two fast waves; and
iii) the interaction between a fast wave and an Alfv\'en wave.  

Basic constraints on the mode interactions arise from conservation
of energy and momentum.  Because the relativistic magnetofluid is
inherently compressible, there is guaranteed to be a three-mode
coupling between two Alfv\'en waves and a fast wave
\cite{TB98}.  The Alfv\'en waves ($A1$, $A2$) satisfy the dispersion
relation $\omega_A = \pm k^z_A$; and the fast mode $\omega_F = 
|{\bf k}_F|$.  Conservation of energy
\be\label{frecon}
\omega_{A1} + \omega_{A2} = \omega_F
\ee
and longitudinal momentum
\ba\label{kcon}
k^z_{A1} + k^z_{A2} &=& \omega_{A1} - \omega_{A2}\nn
 &=&k^z_F = \omega_F\cos\theta_F
\ea
($\cos\theta_F = {\bf k}_F\cdot{\bf B}_0/k_FB_0$) can be combined to give
\be
\cos\theta_F = {\omega_{A1}-\omega_{A2}\over\omega_{A1}+\omega_{A2}}.
\ee
Thus in a frame in which the two colliding Alfv\'en modes have
equal frequencies, the fast mode is emitted perpendicular
to the background field and with a frequency $\omega_F = 2\omega_A$ -- 
just as we found in Section \ref{acol}.

There is also a three-mode interaction between the two colliding 
Alfv\'en waves and a third Alfv\'en wave.  From a kinematic viewpoint, this
interaction is non-vanishing only if at least one of the colliding
modes has a zero-frequency component (\cite{Mont95}, \cite{Ng96}),
as is easily checked by
setting $\cos\theta_F = 1$ in Eqs. (\ref{frecon}) and (\ref{kcon}).
An alternative description is that the magnetic field lines
experience a net displacement (or braiding) across at least one
of the colliding wavepackets \cite{GS97}, \cite{TB98}.  
(A further requirement for a non-vanishing three-mode interaction,
as we detail shortly, is that the colliding modes are non-axisymmetric.)

Next consider the interaction between two fast waves.  It is here
that a significant difference arises between these two viewpoints.
If the background magnetic field suffers a net displacement
across the fast wavefront, then a zero-frequency component
is present in the fourier decomposition of the field.
From a kinematic viewpoint, this zero-frequency component would
facilitate a three-wave interaction between the two colliding fast
modes and a third fast mode.
Two colliding fast waves of finite frequency can generate
a third fast wave only if the waves are colinear:  only in that
case is it possible to satisfy conservation of energy
\be
\omega_{F1} + \omega_{F2} = \omega_{F3}
\ee
and momentum
\be
{\bf k}_{F1} + {\bf k}_{F2} = {\bf k}_{F3}.
\ee
(Two photons propagating obliquely in vacuum
cannot merge to form a single photon, in part 
because these kinematic conditions
cannot be satisfied.)  
From the same kinematic viewpoint a three-wave interaction 
between two colliding fast waves and an Alfv\'en mode is possible 
only in the degenerate case where both colliding waves propagate
along the background magnetic field.\footnote{We thank Maxim Lyutikov
for an illuminating discussion of this point.} This restriction
would appear to be somewhat relaxed if one of the waves has a zero-frequency
component:  then the direction of propagation of that wave can
be arbitrary.

As our second task in this
section, we calculate the form of this three-mode interaction 
for two colliding planar fast waves.  We find that, in fact,
no fast wave is emitted during the collision,
but an Alfv\'en wave is.   The Alfv\'en wave is emitted for
any direction of propagation of the two colliding waves.
Moreover, when there is a field line
displacement in one fast wave, and the other is a pure sinusoid,
then the outgoing Alfv\'en wave is also a pure sinusoid but
its frequency is not obtainable from a three-wave resonance condition.
It is only when the two colliding modes are directed along the
background magnetic field that the resonance condition can
be satisfied with a zero-frequency component.

The fast mode also undergoes a four-mode interaction analogous
to photon scattering, with the conservation relations
\be
\omega_{F1} + \omega_{F2} = \omega_{F3} + \omega_{F4}
\ee
and 
\be
{\bf k}_{F1} + {\bf k}_{F2} = {\bf k}_{F3} + {\bf k}_{F4}.
\ee
However, the fast mode also has a non-linear interaction with
an Alfv\'en mode.  We show that if the Alfv\'en wave spectrum
is that of a critically balanced cascade (\cite{gs96}), then
this second interaction dominates the self-interaction of the fast
mode (as was claimed without detailed justification in \cite{TB98}).

\subsection{Collision between two Alfv\'en waves}

A collision between two
axisymmetric torsional modes generates a compressive motion
transverse to the axis of the background magnetic field.  The
amplitude of this compressive motion is, however, strongly
suppressed when the colliding wavepackets are highly elongated
($k_zR_{\rm max} \ll 1$).  In the case of two monochromatic
torsional waves with frequencies $\omega = \mp k_z$,
this radial disturbance vanishes at the outer boundary of the
zone supporting the colliding Alfv\'en waves.  More generally, 
$\delta R_{\rm max}$ is suppressed by a factor $\sim (k_zR_{\rm max})^2$, 
compared with the result of simple dimensional analysis,
\be
{\delta R_{\rm max}\over R_{\rm max}}
\sim (k_zR_{\rm max})^2 \times R_{\rm max}^2
{\partial\phi_+\over\partial z}{\partial\phi_-\over\partial z}.
\ee
In other words, the three-mode interaction between two Alfv\'en waves
and the fast mode is suppressed for elongated wavepackets
(such as are created during a weak turbulent cascade \cite{gs96}).

There is, nonetheless, a much stronger three-mode interaction involving
a third Alfv\'en wave.  This interaction is not apparent if
one restricts the two colliding waves to be axially symmetric,
and so we will broaden the analysis here to include Alfv\'en modes
supported by coordinate perturbations $\delta\phi$ and $\delta r$.
The relevant component of the force-free equation is
\be\label{ffmink}
J^{t(1)}F_{t\phi}^{(1)} + J^{z(1)}F_{z\phi}^{(1)}
+ J^{r(2)}F_{r\phi}^{(0)} = 0.
\ee
In Minkowski space, $\delta F_{t\phi}^{(1)} = (B_0r) \partial_t\delta r$
and $\delta F_{z\phi}^{(1)} = (B_0r) \partial_z\delta r$.  
The second-order current vanishes for the two separate modes $+$ and $-$,
but where they overlap the first two terms in Eq. (\ref{ffmink})
do not cancel:
\be\label{twoterms}
J^{t(1)}F_{t\phi}^{(1)} + J^{z(1)}F_{z\phi}^{(1)}
= {B_0^2\over 2\pi}\left[
\partial_\phi\left(\partial_z\delta r_+\partial_z\delta r_-\right)
-\partial_z\delta r_+\,\partial_r\left(r^2\partial_z\delta\phi_-\right)
- \partial_z\delta r_-\,\partial_r\left(r^2\partial_z\delta\phi_+\right)
\right].
\ee

When at least one of the colliding Alfv\'en wavepackets
is not axisymmetric (with both perturbations $\delta\phi$ and $\delta r$
excited), one also finds an explicit second order contribution to the
fields,
\be
\delta F^{(2)}_{rt} = 
\left[{1\over r}\partial_r\left(r \delta r_+\right)\partial_t\delta\phi_-
+ {1\over r}\partial_r\left(r \delta r_-\right)\partial_t\delta\phi_+
-\partial_t\delta r_+ \partial_r\delta\phi_-
-\partial_t\delta r_- \partial_r\delta\phi_+\right]B_0 r;
\ee
and similarly for $\delta F^{(2)}_{r\phi}$ and $\delta F^{(2)}_{r z}$.
The second-order current therefore has an explicit contribution
from the colliding modes.
As before, an additional interaction component 
$\delta r_I$, $\delta\phi_I$ to the coordinate fields is 
required to solve the force-free equation:
\ba
4\pi J^{r(2)} &=& -\partial_t\delta F^{(2)}_{rt} +
\partial_z\delta F^{(2)}_{rz} + 
{1\over r^2}\partial_\phi\delta F^{(2)}_{r\phi}\nn
&=& -2B_0r\Bigl[\partial_z\delta r_+\partial_r\partial_z\delta\phi_-
+\partial_z\delta r_-\partial_r\partial_z\delta\phi_+
+\partial_\phi\left(\partial_z\delta\phi_+\partial_z\delta\phi_-\right)\Bigr]
+{1\over r^2}\partial_\phi\delta F^{(2)}_{r\phi}
\nn
&+& B_0r\left\{-\partial_t^2\delta\phi_I +
\partial_z^2\delta\phi_I + \partial_\phi\left[{1\over r^3}
\partial_r\left(r\delta r_I\right) + 
{1\over r^2}\partial_\phi\delta\phi_I\right]
\right\}.
\ea
Thus Eqs.  (\ref{ffmink}) and (\ref{twoterms}) combine to
give the equation of motion
\ba\label{aaaint}
-\partial_t^2\delta\phi_I +
\partial_z^2\delta\phi_I &+& {1\over r^2}\partial_\phi\left[{1\over r}
\partial_r\left(r\delta r_I\right) + \partial_\phi\delta\phi_I
+ {1\over B_0r}\delta F^{(2)}_{r\phi}
\right]
\nn
&=& {2\over r}\left[
2\partial_z\delta r_+\,\partial_r\left(r\partial_z\delta\phi_-\right)
+ 2\partial_z\delta r_-\,\partial_r\left(r\partial_z\delta\phi_+\right)
+ r\partial_\phi\left(\partial_z\delta \phi_+\partial_z\delta \phi_-\right)
- {1\over r}\partial_\phi\left(\partial_z\delta r_+\partial_z\delta r_-\right)
\right].\nn
\ea

In this equation, the transverse components of the laplacian on
the left-hand side suppress the interaction unless they vanish --
that is, unless the the new mode is an Alfv\'en wave (Eq. \ref{alfcon}).
Even without making this restriction, the dominant three-mode interaction
can be obtained by transforming to the variable
\be
\Gamma_\pm \equiv \oint d\phi\;\delta\phi_\pm(\phi, r);\;\;\;\;\;\;
\Gamma_I \equiv \oint d\phi\;\delta\phi_I(\phi, r).
\ee
Here, the integral is performed at constant cylindrical radius.
Applying the operator $\oint d\phi$ to Eq. (\ref{aaaint})
kills off the last term on each side, and we are left with
\be
-\partial_t^2\Gamma_I + \partial_z^2\Gamma_I = 
{4\over r}\oint d\phi\left[
\partial_z\delta r_+\,\partial_r\left(r\partial_z\delta\phi_-\right)
+ \partial_z\delta r_-\,\partial_r\left(r\partial_z\delta\phi_+\right)
\right].
\ee
Transforming to light-cone variables $z_\pm = z \pm t$,
$\partial_\pm = {1\over 2}(\partial_z\pm\partial_t)$, the three-mode
correction to each Alfv\'en mode can be calculated from
\ba
\partial_-\Gamma_+|_{\rm out} - \partial_-\Gamma_+|_{\rm in} 
  &=&  \int d z_+ \partial_+\partial_-\Gamma_I\nn
  &=&  {1\over r}\oint d\phi\left[
\partial_z\delta r_+\,\partial_r\left(r\partial_z\Delta\phi_-\right)
+ \partial_z\Delta r_-\,\partial_r\left(r\partial_z\delta\phi_+\right)
\right].
\ea
The correction to $\Gamma_-$ is obtained by interchanging $+$ and $-$ in
this equation.

As advertised, the resonant three-mode interaction
depends on the presence of a net 
twist (radial displacement) of the magnetofluid across the wavepacket,
\be
\Delta\phi_\pm = \int dz_\mp\;\partial_z\delta\phi_\pm(z_\mp);\;\;\;\;\;\;
\Delta r_\pm = \int dz_\mp\;\partial_z\delta r_\pm(z_\mp).
\ee
Because the colliding modes are Alfv\'en modes,
$\Delta r_\pm$ is related to $\Delta\phi_\pm$ through
the condition of incompressibility,
\be
{1\over r}\partial_r\left(r\Delta r_\pm\right) + 
\partial_\phi\Delta\phi_\pm = 0.
\ee
The three-mode interaction also depends essentially
on the non-axisymmetry of the colliding wavepackets (i.e., on the 
simultaneous presence of coordinate perturbations
$\delta r$ and $\delta\phi$).  Finally we note that applying the
operator $\int d\phi$ to the constraint equation (\ref{fastcon})
for the fast mode gives
\be
{\partial\over\partial r}\left(r^2\int d\phi \delta\phi\right) = 0,
\ee
hence $r^2\Gamma =$ constant.  As long
as $\Gamma$ does not have a singularity at $r = 0$, we see 
that $\Gamma = 0$ for the fast mode. 

\subsection{Collision between two fast waves}

In order to tackle the collision between two fast waves, we consider
the special case of two planar waves.   (There are inherent complications
associated with the choice of cylindrical geometry, due to the 
the non-locality of waveforms propagating in two spatial dimensions --
in contrast with one or three dimensions, the Greens function is
not a delta function.)

Choosing a background magnetic field ${\bf B}_0 = B_0\hat z$, the
perturbed Faraday tensor is
\be
F_{\mu\nu} = \left(\partial_\mu x^0\partial_\nu y^0-\partial_\mu y^0
\partial_\nu x^0\right)B_0.
\ee
Each wave, by itself, is equivalent to a plane electromagnetic wave
superposed on a uniform background magnetic field, and the associated
coordinate perturbation satisfies the usual wave equation,
\be
x^0 = x + \delta x^0_{1,2}(x^\mu);\;\;\;\;\;\;
\partial_\nu\partial^\nu\left[\delta x^0_{1,2}(x^\mu)\right] = 0.
\ee
We can boost along the background magnetic field into a frame 
in which one of the colliding fast waves propagates in a direction 
perpendicular to ${\bf B}_0$, say
\be
\delta{\bf x}^0_1 = \delta x^0_1(x-t)\hat x.
\ee
A specific example of such a waveform is the harmonic perturbation
${\bf k}_1 = k_1 \hat x$, $\delta x^0_1 = \delta X_1 e^{ik_1(x-t)}$.
Because neither fast mode involves a propagating current in the force-free
limit, the interaction
vanishes if the electric vectors of the two modes are parallel and
their direct superposition involves no violation of the constraint
$\delta{\bf E}\cdot{\bf B}_0 = 0$.  Thus the waveform of the
second mode is chosen to have a non-vanishing derivative in the
$y$-direction.  (In general all three components of ${\bf k}_2$ 
are non-vanishing.)  The corresponding coordinate fluctuation is
\be
\delta{\bf x}^0_2 = \delta y^0_2({\bf n}_2\cdot{\bf x}-t)
\left[\hat y + {n_{2x}\over n_{2y}}\hat x\right],
\ee
where ${\bf n}_2$ is an arbitrary unit vector. 
The relative normalization of the fluctuations in $x^0$ and $y^0$ is
fixed by the vanishing of the current,
$\partial_y\delta x^0_2 - \partial_x\delta y^0_2 = 0$.
We allow for the possibility that there is a non-vanishing
displacement across (at least) one of the colliding waveforms, e.g.
\be\label{dispx}
\Delta x^0_1 = \int d\ell (\delta x^0_1)'(\ell) \neq 0;\;\;\;\;\;\;
\Delta y^0_2 = \int d\ell (\delta y^0_2)'(\ell) \neq 0.
\ee

The current obtained by superposing $\delta{\bf x}^0 = \delta{\bf x}^0_1 +
\delta{\bf x}^0_2$ is 
\be\label{jintf}
4\pi J^\mu_{FF} = 
\partial_\nu\left(\partial^\mu x^0\partial^\nu y^0-\partial^\mu y^0
\partial^\nu x^0\right)B_0
   = \left(\partial_\nu\partial^\mu \delta x^0_1 \partial^\nu \delta y^0_2
    - \partial_\nu\partial^\mu \delta y^0_2 \partial^\nu \delta x^0_1\right)B_0.
\ee
Because the linear currents $J^{\mu(1)}$ vanish, the force-free condition
is $J^{x(2)} = J^{y(2)} = 0$ to second order in the wave amplitude.
We must then introduce an interaction field ${x^\mu}_I^0$ and
associated current
\be
4\pi J^\mu_I = \left[\partial_y\partial^\mu x_I^0 - 
     \partial_x\partial^\mu y^0_I
   + \delta^\mu_x \left(\partial_\nu\partial^\nu y^0_I\right) - 
     \delta^\mu_y \left(\partial_\nu\partial^\nu x^0_I\right)\right]B_0.
\ee
The conditions
\be
J^x_I + J^x_{FF} = 0 =  J^y_I + J^y_{FF}
\ee
correspond to the equations of motion
\ba
\partial_\nu\partial^\nu y^0_I &+&
\partial_x(\partial_y\delta x^0_I - \partial_x\delta y^0_I) =\cr
&&(1-n_{2x})\left[(\delta x^0_1)''(x-t)\cdot 
(\delta y^0_2)'({\bf n}_2\cdot{\bf x}-t)  -
n_{2x}(\delta x^0_1)'(x-t)\cdot 
(\delta y^0_2)''({\bf n}_2\cdot{\bf x}-t)\right]\nn
\ea
and
\be
\partial_\nu\partial^\nu x^0_I 
- \partial_y(\partial_y\delta x^0_I - \partial_x\delta y^0_I) = 
n_{2y}(1-n_{2x})\cdot(\delta x^0_1)'(x-t)\cdot 
(\delta y^0_2)''({\bf n}_2\cdot{\bf x}-t).
\ee
It is useful to project these equations onto the outgoing fast 
($\partial_x x^0_I + \partial_y y^0_I$) and Alfv\'en 
($\partial_x y^0_I - \partial_x y^0_I$) modes,
\be\label{fasteq}
\partial_\nu\partial^\nu\left(\partial_x x^0_I + \partial_y y^0_I\right)
= 2(1-n_{2x})n_{2y} (\delta x^0_1)''(x-t)\cdot 
(\delta y^0_2)''({\bf n}_2\cdot{\bf x}-t)
\ee
and
\ba\label{alfeq}
(-\partial_t^2+\partial_z^2)\left(\partial_x y^0_I - \partial_y x^0_I\right)
= (1-n_{2x})&\Bigl[&(\delta x^0_1)'''(x-t)\cdot
(\delta y^0_2)'({\bf n}_2\cdot{\bf x}-t) -\nn
&&(n_{2x}^2+n_{2y}^2)(\delta x^0_1)'(x-t)\cdot 
(\delta y^0_2)'''({\bf n}_2\cdot{\bf x}-t)\;\Bigr].\nn
\ea

To find the outgoing perturbation to the fast wave 1, we 
start with the Greens function solution to Eq. (\ref{fasteq}),
\be
\left(\partial_x x^0_I+\partial_y y^0_I\right)({\bf x},t) = 
2n_{2y}(1-n_{2x})\int d^3x' {1\over|{\bf x}-{\bf x'}|}
(\delta x^0_1)''(x'-t+|{\bf x}-{\bf x}'|)\cdot
(\delta y^0_2)''({\bf n}_2\cdot{\bf x}' - t +|{\bf x}-{\bf x}'|),
\ee
(using the retarded time $t - |{\bf x}-{\bf x}'|$)
and take the limit $x,t\rightarrow \infty$ at $y=z=0$.
Each wavepacket is localized within a distance $\sim L$ along
its direction of propagation.
Transforming $x' \rightarrow x'-x$ and then setting
$\{x',y',z'\} = r\{\sin\theta\cos\phi,\sin\theta\sin\phi,\cos\theta\}$,
one has
\ba\label{fasteqb}
\left(\partial_x x^0_I+\partial_y y^0_I\right)(x,0,0,t) 
= 2n_{2y}(1-n_{2x})\int rdr\sin\theta d\theta d\phi\;
\Bigl\{\,(\delta x^0_1)''[r(1+\sin\theta\cos\phi) + x-t]\cr
\;(\delta y^0_2)''[r(1+{\bf n}_2\cdot{\bf\Omega}) + n_{2x}x -t]\,\Bigr\}.\cr
\ea
The integrand is non-vanishing only in 
a small region $\Delta r \sim L$ near 
$r = (t~-~n_{2x}x)/(1~+~{\bf n}_2\cdot{\bf\Omega})$ and
$\Delta\theta\Delta\phi \sim L/t$ near
$\theta = \pi/2$, $\phi = \pi$.  
When integrating over $r$, the argument of $(\delta x^0_1)''$ varies only
by a small amount $\sim L^2/t$.  The integral 
$\int d\ell (\delta y^0_2)''(\ell)$ itself vanishes, and so the 
outgoing
fast mode has an amplitude $\sim (L/t)(\delta x^0_1)'(\delta y^0_2)' 
\rightarrow 0$ as $t \rightarrow \infty$.   We conclude that the two
fast modes do {\it not} couple to a third fast mode, even if one of
the colliding waveforms has a non-vanishing displacement 
(\ref{dispx}).

Now let us examine the coupling to an outgoing Alfv\'en mode.  
Transforming to the light-cone coordinates $z_\pm = z\pm t$,
and taking $x=y=0$, Eq. (\ref{alfeq}) becomes
\ba
\partial_+\partial_-\Bigl(\partial_x y^0_I&-&\partial_y x^0_I\Bigr)
(0,0,z_+,z_-) 
=\nn
&&{1\over 4}(1-n_{2x})
\biggl\{(\delta x^0_1)'''\left[{1\over 2}(z_- - z_+)\right]\cdot
(\delta y^0_2)'\left[{1\over 2}(n_{2z}+1)z_- + {1\over 2}(n_{2z}-1)z_+\right]
\nn
&-&(n_{2x}^2+n_{2y}^2)\,(\delta x^0_1)'\left[{1\over 2}(z_- - z_+)\right]\cdot
(\delta y^0_2)'''\left[{1\over 2}(n_{2z}+1)z_- + 
{1\over 2}(n_{2z}-1)z_+\right]\biggr\}.\nn
\ea
For example, the outgoing Alfv\'en wave propagating to large positive $z$ is
\ba\label{alfout}
\partial_-\Bigl(\partial_x y^0_I&-&\partial_y x^0_I\Bigr)
(0,0,z-t) = \nn
&&{1\over 4}(1-n_{2x})\int dz_+
\biggl\{(\delta x^0_1)'''\left[{1\over 2}(z_- - z_+)\right]\cdot
(\delta y^0_2)'\left[{1\over 2}(n_{2z}+1)z_- + {1\over 2}(n_{2z}-1)z_+\right]
\nn
&-&(n_{2x}^2+n_{2y}^2)\,(\delta x^0_1)'\left[{1\over 2}(z_- - z_+)\right]\cdot
(\delta y^0_2)'''\left[{1\over 2}(n_{2z}+1)z_- + 
{1\over 2}(n_{2z}-1)z_+\right]\biggr\}.\nn
\ea
This integral has a closed form solution when one of the wavepackets
(say 2) is much shorter than the other and has a zero-frequency component.
Then we can treat $(\delta y^0_2)'$ as a delta function,
$(\delta y^0_2)'(\ell) = \Delta y^0_2 \delta(\ell)$.  
After restoring the dependence on the transverse coordinates $x$ and $y$, 
and integrating once over $z_-$, our final expression for the outgoing wave is
\be
\Bigl(\partial_x y^0_I-\partial_y x^0_I\Bigr)(x,y,z-t) = 
{(1-n_{2x})\over 2n_{2z}}\Delta y^0_2\cdot(\delta x^0_1)''
\left[{1-n_{2x}-n_{2z}\over 1-n_{2z}}x - {n_{2y}\over 1-n_{2z}}y
-{n_{2z}\over 1-n_{2z}}(z-t)\right].
\ee
The outgoing Alfv\'en wave is sheared, with wavevector proportional to
\be
(\omega,{\bf k})
\propto \left(n_{2z},n_{2x}+n_{2z}-1, n_{2y}, n_{2z}\right).
\ee
Notice also that the amplitude of the outgoing wave involves a higher-order
derivative of the ingoing wave, and so the higher-wavenumber fourier
components are enhanced by the collision.  

The frequency of the outgoing Alfv\'en wave is 
\be\label{omegaa}
\omega_A = {n_{2z}-n_{1z}\over n_{2z}-1}\omega_{F1}
\ee
when the fast wave 1 is a pure fourier mode with frequency $\omega_{F1}$,
and its wavevector has some component along the background magnetic field.
The three-wave resonance condition
\be
\omega_A = \omega_{F1}+\omega_{F2} = \omega_{F1};\;\;\;\;\;\;
k^z_A = k^z_{F1} + k^z_{F2} = k^z_{F1},
\ee
can be satisfied with $\omega_{F2} = k^z_{F2} = 0$
if the propagation of fast wave 1 is aligned 
with the background field ($n_{1z} = 1$).
But, more generally, Eq. (\ref{omegaa}) is not consistent with such a 
three-wave resonance:  the integral over $z_+$ in Eq. (\ref{alfout}) 
is non-vanishing only if 
\be
(n_{1z}-1)\omega_{F1} + (n_{2z}-1)\omega_{F2} = 0.
\ee
A solution to this equation, with at least one non-vanishing frequency, 
requires that either $n_{1z} = 1$ or $n_{2z} = 1$.  This illustrates
why a net field line displacement across an MHD wave has only a
limited description in terms of a zero-frequency component of the wave.

\subsection{Collision between a fast wave and an Alfv\'en wave}

The collision between a fast wave and an Alfv\'en wave can generate
both an outgoing Alfv\'en mode and an outgoing fast mode through
three-wave interactions.  It is useful to compare the strength
of this interaction with the interaction between two
fast modes discussed in the previous section.  We focus on
the particular case where the Alfv\'en wave spectrum
is highly anisotropic, and the coupling parameter $k_yx^0 - k_xy^0
\sim (k_\perp/\omega)(\delta B_{A,\omega}/B)$ is independent of scale and 
close to unity.  A key difference between these two types of
wave collisions is that the sheared Alfv\'en waves are 
current-carrying, $J^{(1)}_{A,\omega} \sim k_\perp \delta B_{A,\omega}$ 
and there is a second-order Lorentz force involving
the interaction between this current and the fluctuating field of
the fast wave.  The magnitude of this Lorentz force is
\be
4\pi J^{(1)}_{A,\omega} \delta B_{F,\omega}
\sim k_\perp \delta B_{A,\omega} \delta B_{F,\omega}.
\ee
Here $\delta B_{A,\omega}$ and $\delta B_{F,\omega}$ are the field 
perturbations of the Alfv\'en and fast waves at frequency $\omega$.
We can compare this with the second order current generated in the 
fast-wave collision,
\be
4\pi J^{(2)}_{FF} \sim \omega {\delta B_{F,\omega}^2\over B_0}.
\ee
The ratio is
\be
{J^{(1)}_{A,\omega} \delta B_{F,\omega}\over J^{(2)}_{FF}B_0}
\sim \left({k_\perp \delta B_{A,\omega}\over \omega B_0}\right)
\left({\delta B_{F,\omega}\over B_0}\right)^{-1}
\sim \left({\delta B_{F,\omega}\over B_0}\right)^{-1} \gg 1.
\ee
Thus, if the Alfv\'en waves form an anisotropic cascade, then the
fast wave spectrum evolves predominantly through collisions with
Alfv\'en waves and not through self-collisions between fast waves.

\begin{acknowledgments}
We thank the NSERC of Canada for its financial support, and Maxim
Lyutikov for conversations.
\end{acknowledgments}

\appendix
\section{}\label{appone}

In this appendix, we show how the force-free equation
may be derived from the action
\be\label{smag}
S' = \int d^4x_0 L^\prime = 
\int d^4x \frac {1}{4} \sqrt{-g} g_{\mu\rho} g_{\nu\sigma}
{\widetilde F}^{\mu\nu}{\widetilde F}^{\rho\sigma} =
\int d^4x_0 \frac{J_4}{4} \sqrt{-g} g_{\mu\rho} g_{\nu\sigma}
{\widetilde F}^{\mu\nu} {\widetilde F}^{\rho\sigma},
\ee
using the magnetic lagrangian fluid variables (\ref{magvar}).
Varying Eq. (\ref{smag}),
\be
\delta S' = \int d^4x_0 \delta L^\prime  = 
\int d^4x_0\left [ \frac{\partial L^\prime}{\partial x^{\mu}} \delta x^{\mu} + 
\frac{\partial L^\prime}{\partial (\partial x^{\mu}/{x_0}^{\alpha})}\,
\delta\biggl(\frac{\partial x^{\mu}}{\partial {x_o}^{\alpha}}\biggr)\right],
\ee
there are terms which arise from the explicit
$x^\mu$-dependence of the metric,
\be\label{parone}
\frac{\partial L^\prime}{\partial x^{\mu}} = 
\left[\frac{(-g_0)}{4J_4} 
{\partial x^\beta\over\partial x_0^\kappa}
{\partial x^\gamma\over\partial x_0^\lambda}
{\partial x^\delta\over\partial x_0^\eta}
{\partial x^\varepsilon\over\partial x_0^\zeta}
{\widetilde F_0}^{\kappa\lambda} {\widetilde F_0}^{\eta\zeta}\right]
\frac{\partial}{\partial x^\mu} \left[
{g_{\beta\delta} g_{\gamma\varepsilon}\over \sqrt{-g}}\right],
\ee
as well as from the dependence of $\widetilde F^{\mu\nu}$ on
$\partial x^\mu/\partial x_0^\alpha$,
\be\label{partwo}
\frac{\partial L^\prime}{\partial (\partial x^{\mu}/{x_0}^{\alpha})} =
\left[{g_{\beta\delta} g_{\gamma\varepsilon}\over\sqrt{-g}}\right]
\frac{\partial}{\partial (\partial x^{\mu}/{x_0}^{\alpha})} 
\left[\frac{(-g_0)}{4J_4} 
{\partial x^\beta\over\partial x_0^\kappa}
{\partial x^\gamma\over\partial x_0^\lambda}
{\partial x^\delta\over\partial x_0^\eta}
{\partial x^\varepsilon\over\partial x_0^\zeta}
{\widetilde F_0}^{\kappa\lambda} {\widetilde F_0}^{\eta\zeta}\right].
\ee

After extremizing, the usual Euler-Lagrange equations are obtained,
\be
\frac {\partial}{\partial x_0^\alpha}
\left[\frac {\partial L^\prime} 
{\partial (\partial x^{\mu}/\partial x_0^\alpha)}\right]
- \frac {\partial L^\prime}{\partial x^{\mu}} = 0.
\ee
Substituting eqs. (\ref{parone}) and (\ref{partwo}) then gives
\ba\label{ffb}
{\partial\over\partial x_0^\alpha}
\Biggl[g_{\beta\delta}g_{\gamma\mu}\widetilde F^{\beta\gamma}
\left(\sqrt{-g_0}{\partial x^\delta\over\partial x_0^\eta}
\widetilde F_0^{\eta\alpha}\right)
- {\partial J_4\over\partial(\partial x^\mu/
\partial x_0^\alpha)} {\sqrt{-g}\over 4}g_{\beta\delta}g_{\gamma\varepsilon}
\widetilde F^{\beta\gamma} \widetilde F^{\delta\varepsilon}\Biggr]\nn
-{(-g)J_4\over 4}\widetilde F^{\beta\gamma} 
\widetilde F^{\delta\varepsilon}{\partial\over\partial x^\mu}
\left[{g_{\beta\delta}g_{\gamma\varepsilon}\over\sqrt{-g}}\right] = 0. 
\ea
Now the background field satisfies
\be
\frac{\partial}{\partial x_0^\alpha}
\left(\sqrt{-g_0}{\partial x^\delta\over\partial x_0^\eta}
\widetilde F_0^{\eta\alpha}\right) = 0.
\ee
Making use of this relation, and the identities
\be
\frac{\partial}{\partial x_0^\alpha}
\left[\frac{\partial {J_4}}
{\partial (\partial x^\mu/\partial x_0^\alpha)}\right] = 0;
\;\;\;\;\;\;J_4{\partial\over\partial x^\mu} =
\frac{\partial {J_4}}
{\partial (\partial x^\mu/\partial x_0^\alpha)} 
{\partial\over \partial x_0^\alpha}
\ee
in Eq. (\ref{ffb}) gives 
\be
\widetilde F^{\delta\varepsilon}{\partial\over\partial x^\varepsilon}
\left[g_{\beta\delta}g_{\gamma\mu}\widetilde F^{\beta\gamma}\right]
- {1\over 2}\widetilde F^{\delta\varepsilon}
{\partial\over\partial x^\mu}\left[g_{\beta\delta}g_{\gamma\varepsilon}
\widetilde F^{\beta\gamma}\right] = 0.
\ee

To demonstrate the equivalence of this expression and the usual
force free equation, one need only substitute
$F_{\alpha\beta} = -{1\over 2}\sqrt{-g}\varepsilon_{\alpha\beta\rho\sigma}
\widetilde F^{\rho\sigma}$ in 
\be
J^\mu F_{\mu\nu} = {1\over\sqrt{-g}}{\partial\over\partial\rho}
\left(\sqrt{-g}g^{\mu\alpha}g^{\rho\beta}F_{\alpha\beta}\right)F_{\mu\nu} = 0.
\ee
After making use of the contraction
\be g^{\mu\alpha}\varepsilon_{\alpha\beta\gamma\delta}
\varepsilon_{\mu\nu\rho\sigma} = 
-4{g_{\beta\nu}g_{\gamma\rho}g_{\delta\sigma}\over (-g)}
\left[\delta_\beta^\varepsilon \delta_\gamma^\kappa \delta_\delta^\lambda
- \delta_\beta^\varepsilon \delta_\delta^\kappa \delta_\gamma^\lambda
+ \delta_\delta^\varepsilon \delta_\beta^\kappa \delta_\gamma^\lambda
- \delta_\delta^\varepsilon \delta_\gamma^\kappa \delta_\beta^\lambda
+ \delta_\gamma^\varepsilon \delta_\delta^\kappa \delta_\beta^\lambda
- \delta_\gamma^\varepsilon \delta_\beta^\kappa \delta_\delta^\lambda\right],
\ee
Eq. (\ref{ffc}) is obtained.

\section{}\label{apptwo}

In this appendix we list, for reference, the axisymmetric perturbations of
a background magnetofluid satisfying $J^\mu = 0$, using the
electric Lagrangian fluid variables (\ref{transf}).

The field strengths are given by
\be
F_{r \phi} =
\left(\frac {\partial r_0}{\partial r}
\frac {\partial {\phi}_0}{\partial \phi} -
\frac {\partial {\phi}_0}{\partial r}
\frac {\partial r_0}{\partial \phi}\right) F^{(0)}_{r \phi} +
\left(\frac {\partial r_0}{\partial r}
\frac {\partial {t}_0}{\partial \phi} -
\frac {\partial {t}_0}{\partial r}
\frac {\partial r_0}{\partial \phi}\right) F^{(0)}_{rt},
\ee
where the background fields $F^{(0)}_{r \phi}$ and $F^{(0)}_{rt}$ are 
\be
F^{(0)}_{r \phi} = \frac {B_0 g_{0\,\phi\phi} g_{0\,rr}}{\sqrt{-g_0} R_0};
\ee
\be
F^{(0)}_{rt} = - \left({g_0^{t \phi} \over g_0^{tt}}\right) 
F^{(0)}_{r \phi}.
\ee

In a cylindrically symmetric spacetime with $z_0 = z$,
$\partial\phi_0/\partial\phi = 1$, and $\partial r_0/\partial\phi = 
\partial t_0/\partial\phi = 0$,
the perturbations  $\delta F_{\mu \nu} = F_{\mu \nu} -  F^{(0)}_{\mu \nu}$ 
have the form 
\be
\delta F_{r \phi} =
\left(\frac {\partial r_0}{\partial r}\right)F^{(0)}_{r \phi}(r_0) - 
F^{(0)}_{r \phi}(r);
\ee
\be
\delta F_{r t} =
\left(\frac {\partial r_0}{\partial r}
\frac {\partial {\phi}_0}{\partial t} -
\frac {\partial {\phi}_0}{\partial r}
\frac {\partial r_0}{\partial t}\right) F^{(0)}_{r \phi}(r_0) +
\left(\frac {\partial r_0}{\partial r}
\frac {\partial {t}_0}{\partial t} -
\frac {\partial {t}_0}{\partial r}
\frac {\partial r_0}{\partial t}\right) F^{(0)}_{rt}(r_0) - F^{(0)}_{rt}(r);
\ee
\be
\delta F_{r z} =
\left(\frac {\partial r_0}{\partial r}
\frac {\partial {\phi}_0}{\partial z} -
\frac {\partial {\phi}_0}{\partial r}
\frac {\partial r_0}{\partial z}\right) F^{(0)}_{r \phi}(r_0) +
\left(\frac {\partial r_0}{\partial r}
\frac {\partial {t}_0}{\partial z} -
\frac {\partial {t}_0}{\partial r}
\frac {\partial r_0}{\partial z}\right) F^{(0)}_{rt}(r_0);
\ee
\be
\delta F_{t \phi} =
\left(\frac {\partial r_0}{\partial t}\right) F^{(0)}_{r \phi}(r_0); 
\ee
\be
\delta F_{z \phi} =
\left(\frac {\partial r_0}{\partial z}\right) F^{(0)}_{r \phi}(r_0);
\ee
and
\be
\delta F_{t z} =
\left(\frac {\partial r_0}{\partial t}
\frac {\partial {\phi}_0}{\partial z} -
\frac {\partial {\phi}_0}{\partial t}
\frac {\partial r_0}{\partial z}\right) F^{(0)}_{r \phi}(r_0) +
\left(\frac {\partial r_0}{\partial t}
\frac {\partial {t}_0}{\partial z} -
\frac {\partial {t}_0}{\partial t}
\frac {\partial r_0}{\partial z}\right) F^{(0)}_{rt}(r_0).
\ee

Substituting 
$F_{rt}^{(0)}(r_0) = - (g^{t\phi}_0/g^{tt}_0)F_{r\phi}^{(0)}(r_0)$,
these equations reduce to
\be
\delta F_{r \phi} =
\frac {\partial r_0}{\partial r}F^{(0)}_{r \phi}(r_0) 
- F^{(0)}_{r \phi}(r)
\ee
\be
\delta F_{r t} =
\left[\left(\frac {\partial r_0}{\partial r}
\frac {\partial {\phi}_0}{\partial t} -
\frac {\partial {\phi}_0}{\partial r}
\frac {\partial r_0}{\partial t}\right) -
\left(\frac {\partial r_0}{\partial r}
\frac {\partial {t}_0}{\partial t} -
\frac {\partial {t}_0}{\partial r}
\frac {\partial r_0}{\partial t}\right) 
\left({g_0^{t \phi}\over g_0^{tt}}\right)\right] F^{(0)}_{r \phi}(r_0) - 
\left({g^{t \phi}\over g^{tt}}\right) F^{(0)}_{r \phi} (r); 
\ee
\be
\delta F_{r z} =
\left[\left(\frac {\partial r_0}{\partial r}
\frac {\partial {\phi}_0}{\partial z} -
\frac {\partial {\phi}_0}{\partial r}
\frac {\partial r_0}{\partial z}\right) -
\left(\frac {\partial r_0}{\partial r}
\frac {\partial {t}_0}{\partial z} -
\frac {\partial {t}_0}{\partial r}
\frac {\partial r_0}{\partial z}\right)
\left({g_0^{t \phi} \over g_0^{tt}}\right)\right] F^{(0)}_{r \phi}(r_0);
\ee
\be
\delta F_{t \phi} =
\left(\frac {\partial r_0}{\partial t}\right) F^{(0)}_{r \phi}(r_0);
\ee
\be
\delta F_{z \phi} =
\left(\frac {\partial r_0}{\partial z}\right) F^{(0)}_{r \phi}(r_0);
\ee
and
\be
\delta F_{t z} =
\left[\left(\frac {\partial r_0}{\partial t}
\frac {\partial {\phi}_0}{\partial z} -
\frac {\partial {\phi}_0}{\partial t}
\frac {\partial r_0}{\partial z}\right) -
\left(\frac {\partial r_0}{\partial t}
\frac {\partial {t}_0}{\partial z} -
\frac {\partial {t}_0}{\partial t}
\frac {\partial r_0}{\partial z}\right)
\left({g_0^{t \phi} \over g_0^{tt}}\right)\right] F^{(0)}_{r \phi}(r_0).
\ee
For example,
\be
\delta F_{r \phi} =
\left[\left(\frac {\partial r_0}{\partial r}\right)
\frac {g_{0\,rr}g_{0\,\phi\phi}}{\sqrt{-g_0}} -
\frac {g_{rr}g_{\phi\phi}}{\sqrt{-g}}\right] B_0.
\ee

\section{}\label{appthree}

This appendix is devoted to a derivation of the force-free equation
(\ref{ffe}) in a non-static, axially symmetric spacetime.
The force-free equation (\ref{ffwj}) is
\ba
\Bigl[{\partial}_{r} \left(\sqrt{-g} g^{tt} g^{rr} \delta F_{tr} +
\sqrt{-g} g^{t \phi} g^{rr} \delta F_{ \phi r}\right) +
{\partial}_{z} \left(\sqrt{-g} g^{t \phi}g^{zz} \delta F_{\phi z}\right)\Bigr] 
F^{(0)}_{tr} &+& \nn
\Bigl[{\partial}_{r}
\left(\sqrt{-g} g^{\phi \phi} g^{rr} \delta F_{\phi r} +
\sqrt{-g} g^{\phi t} g^{rr} \delta F_{tr}\right) +
{\partial}_{z} \left(\sqrt{-g} g^{\phi \phi} g^{zz} \delta F_{\phi z}\right) 
&+& \nn
{\partial}_{t} \left(\sqrt{-g} g^{\phi \phi} g^{tt} \delta F_{\phi t}
+ \sqrt{-g} g^{\phi t} g^{\phi t} \delta F_{t \phi}\right)\Bigr] 
F^{(0)}_{\phi r}= 0.
\ea
where the terms involving $\delta F_{tz}$ have canceled.  
The terms involving $\delta F_{tr}$ may be combined to give
\be\label{Ftrterms}
- \sqrt{-g} g^{tt}g^{rr}
\delta F_{tr}
\frac{\partial}{\partial r}\left(\frac {-g^{t \phi}}{g^{tt}}\right)
F^{(0)}_{r \phi}.
\ee
The terms involving $\delta F_{\phi r}$ combine to give
\be\label{Fphirterms}
\partial_r \left[{\sqrt{-g} \over g_{rr} g_{\phi \phi}}
\frac{\partial}{\partial r}\left(\delta r F^{(0)}_{r \phi}\right)\right] 
F^{(0)}_{r \phi} -
\sqrt{-g} g^{t \phi} g^{rr}
\frac{\partial}{\partial r}
\left(\frac {-g^{t \phi}}{g^{tt}}\right)
\frac{\partial}{\partial r}
\left(\delta r F^{(0)}_{r \phi}\right)F^{(0)}_{r \phi}.
\ee
Here $\det(t,\phi) = g_{tt}g_{\phi\phi} - g_{t\phi}^2$.
Lastly, the terms involving $\delta F_{\phi t}$ and $\delta F_{\phi z}$ 
yield
\be\label{Fphitzterms}
{\sqrt{-g}\over\det(t,\phi)} 
\left(\frac{{\partial}^2 \delta r}{\partial t^2}
+ \frac{g^{zz}}{g^{tt}}
\frac{{\partial}^2 \delta r}{\partial z^2}\right) (F^{(0)}_{r \phi})^2.
\ee

To first order in the coordinate perturbation, the perturbed Maxwell fields 
are
\be
\delta F_{r \phi} = 
\frac{\partial}{\partial r}\left[\delta r F^{(0)}_{r \phi}(r)\right]; 
\ee
\be
\delta F_{rt} = 
\frac{\partial \delta \widetilde{\phi}}{\partial t} F^{(0)}_{r \phi}(r) -
\frac{\partial}{\partial r} \left [ \delta r \frac{g^{t \phi}(r)}{g^{tt}(r)} 
F^{(0)}_{r \phi}(r) \right];
\ee
\be
\delta F_{t \phi} = \frac{\partial \delta r}{\partial t} F^{(0)}_{r \phi}(r);  
\ee
and
\be
\delta F_{z \phi} = \frac{\partial \delta r}{\partial z} F^{(0)}_{r \phi}(r).
\ee
Substituting these expressions and collecting terms, we arrive at
\ba
\frac{1}{\det(t,\phi)} \left[\frac{{\partial}^2 \delta r}{\partial t^2} + 
\frac{g^{zz}}{g^{tt}}
\frac{{\partial}^2 \delta r}{\partial z^2}\right](F^{(0)}_{r \phi})^2 + 
\frac{1}{\sqrt{-g}}\frac{\partial}{\partial r} 
\left[\frac{\sqrt{-g}}{g_{\phi \phi} g_{rr}}
\frac{\partial}{\partial r}
(\delta r F^{(0)}_{r \phi})\right]F^{(0)}_{r \phi}\nn
- g^{tt}g^{rr}\frac{\partial}{\partial r}
\left(\frac {-g^{t \phi}}{g^{tt}}\right)
\left[\frac{{\partial} \widetilde{\phi}}{\partial t} + \delta r 
\frac{\partial}{\partial r}\left(\frac {-g^{t \phi}}{g^{tt}}\right)\right]
(F^{(0)}_{r \phi})^2 &=& 0.
\ea
If we multiply by $\det(\phi,t) g^{tt}$ we obtain equation $(\ref{ffe})$.

\section{}\label{erratum}

This www.arxiv.org manuscript incorporates 
several minor corrections that will appear
as an erratum to the paper as published in the December 2004 
issue of Physical Review D.

1) The second term, RHS, Eq. (100) [and the analogous
term in Eq. (C9)] was spurious and has been deleted.

2) An algebraic error was corrected in the RHS
of Eqns. (181) and (182): the normalization
increased by a factor 2;  and $r^{-4} [r^4] \rightarrow r^{-1} [r]$
in front of [inside] the integral.   The intermediate
equation (179) also received a correction.

3) In Section V.B, a final step has been added
to the derivation of the fast mode amplitude [Eq. (166b)].
The claim [based on Eq. (157)] that  $\delta R_{\rm max}$
can be set to zero was incorrect:  the integral for the
background field energy in Eq. (157) in fact converges
at large radius.  As a result, the LHS of Eq. (166)
gains a term $-{\delta R_{\rm max}/ R_{\rm max}}$,
and the final step Eq. (166b) is required.

%The following (non-propagating) typographical errors 
%were also corrected:

%3) Sign error:  third term, RHS, Eq. (97);  LHS, Eq. 
%(197).

In addition there were a handful of (non-propagating) corrections
to signs and indices.

%5) Spurious factor of $r^{-4}$ in the first term on the RHS
%of Eq. (122) has been deleted.

%6) Indices corrected in $F^{(0)}_{\phi r}$, end of Eq. (C1);
%and in the factor of $(g_{\phi\phi}g_{rr})^{-1}$, LHS
%of Eq. (C3).

\bibliography{ms}

\end{document}